\documentclass[12pt,reqno]{amsart}

\usepackage{graphicx}
\usepackage{amsthm,amsmath,amsfonts,amssymb,amsxtra}

\usepackage{cite}
\usepackage{color}

\newtheorem{theorem}{Theorem}[section]
\newtheorem{lemma}[theorem]{Lemma}

\theoremstyle{definition}

\newtheorem{remark}[theorem]{Remark}

\newtheorem{proposition}[theorem]{Proposition}

\numberwithin{equation}{section}
\newcommand{\T}{{}^{\mathrm t}}

\begin{document}

\title[Antiferromagnetic LRO in a Lattice Fermion Model]{Antiferromagnetic Long-Range Order\\ in a Lattice Fermion Model}


\author[Y. Goto]{Yukimi Goto\textsuperscript{1}}
\thanks{\textsuperscript{1} University of Tokyo,~Graduate School of Mathematical Sciences, Komaba, Meguro-ku Tokyo 153-8914,
	Japan. Email:~{\tt   yukimi@ms.u-tokyo.ac.jp}}

\author[T. Koma]{Tohru Koma\textsuperscript{2}}
\thanks{\textsuperscript{2} Gakushuin University (retired), Department of Physics, Mejiro, Toshima-ku, Tokyo 171-8588, Japan}

\maketitle

\medskip

\noindent
{\bf Abstract:}
We study a lattice fermion model with antiferromagnetic interactions on the three-dimensional cubic lattice.
The hopping term of the Hamiltonian has a Weyl-type dispersion. We prove that the model has reflection positivity.
Moreover, by relying on the property, we prove the existence of the antiferromagnetic long-range order at low temperatures
in a strong coupling regime.

\bigskip

\tableofcontents

\section{Introduction}

The notion of reflection positivity was originally introduced in quantum field theory by Osterwalder and Schrader~\cite{OS},
and it has played an important role in the study of phase transitions for lattice systems, in particular classical and quantum
spin systems, so far.
One of the great successes of the method of reflection positivity is to have proved the existence of long-range order (LRO)
in systems with continuous symmetry~\cite{FSS,DLS,FILS}. However, for fermion systems, the use of the general framework
such as developed in~\cite{FILS1,FILS} was restricted to special fermion systems, such as Majorana fermions~\cite{JP}
and Grassmann fermions in a strong coupling limit \cite{SS1,SS2}.

Quite recently, the method of reflection positivity has been extended to several fermion systems,
which contain superconducting electrons in the $\pi$-magnetic flux \cite{Koma1,Koma2},
and Nambu--Jona-Lasinio models \cite{GK,GK1,GK2} in particle physics.
In this paper, we apply the method of reflection positivity to a lattice Weyl fermion system with antiferromagnetic interactions.
Weyl fermions are attracting considerable interest in modern condensed matter physics because of their topological electronic structure~\cite{VV,AMV,Burkov}.
	The interplay between band topology and magnetism, including antiferromagnetic systems, is reviewed in~\cite{OAKN}.
	These developments motivate the study of lattice fermion models with Weyl-type hopping, and our model provides an idealized setting in which antiferromagnetic long-range order can be established mathematically.

In our previous paper \cite{GK2}, we dealt with the four-component Dirac spinors with Nambu--Jona-Lasinio-type interactions.
In order to realize the reflection positivity, the interactions must be attractive. In the words of condensed matter physics,
the interactions between Dirac spinors are ferromagnetic. In the present paper, we deal with Weyl-type fermions, which have
the usual two-component spin in condensed matter physics.
In the lattice QCD literature, this is also known as the naive fermion, since it is based on a naive discretization~\cite{Rothe}.
Surprisingly, in order to realize the reflection positivity for the Weyl fermions,
the interactions between the spins of the Weyl fermions must be antiferromagnetic (repulsive).
The difference between them comes from the algebraic structures of Dirac $\gamma$ and Pauli matrices.

In fermionic systems, there are very few examples of hopping Hamiltonians that satisfy reflection positivity
in three or higher dimensions, and our example is new as far as we know.
Moreover, we stress that, for a given interacting system, the reflection positivity is non-trivial for the whole Hamiltonian,
even when each of the hopping and the interaction Hamiltonians satisfies reflection positivity independently.

The purpose of our study is to prove the existence of long-range order for the present model at low temperatures in three dimensions
via the reflection positivity for fermion systems inspired by~\cite{JP},
which has been already applied to some models in our earlier works~\cite{GK,GK1,GK2}.
These previous works were motivated by the importance of symmetry breaking in the lattice quantum chromodynamics (QCD) theory
for the study of hadron phenomena.
In contrast to our previous studies, this paper does not address lattice QCD phenomena.
Although our present model is slightly artificial in condensed matter physics,
we believe that our method can be extended to other fermionic systems of interest
in condensed matter physics. For example, see Sec.~8.2 of \cite{Tasaki} about a mathematical approach to
a ferromagnetic long-range order in a $t$-$J$ model,
whose Hamiltonian consists of hopping and interaction terms, and resembles our Hamiltonian in the present paper.
(See also \cite{Tasaki2} and \cite{WY}.)

Our strategy for proving the existence of long-range order is as follows:
We first construct certain unitary transformations so that the Hamiltonian satisfies reflection positivity.
Since we deal with fermions, it is essential to express the hopping Hamiltonian in terms of Majorana fermions~\cite{JP}.
Once the reflection positivity is established, the well-known standard procedure allows us to derive the infrared bound,
which is essentially identical to the corresponding bound in~\cite{DLS, GK}.
To control the Fourier transform of the two-point correlation function, we rely on the methods developed in~\cite{KLS1, KLS2}.

The organization of this paper is as follows.
In Section~\ref{sec.Hm}, we provide the precise definition of the model and the statement of our main result.
In Sections~\ref{sec:Reflec.Preli}--\ref{sec:x2plane}, we establish the reflection positivity
for the model by using the above-mentioned unitary transformations.
In Section~\ref{GdomiInfrab}, we derive the infrared bound from the Gaussian domination bound.
Finally, the proof of LRO at low temperatures is completed by combining the infrared bound
with a certain estimate of the energy expectation value.

\section{Hamiltonian}
\label{sec.Hm}
In order to describe our Hamiltonian, we introduce some notations.
The model is defined on a finite three-dimensional cubic lattice which is given by
$$
\Lambda := \{x = (x^{(1)}, x^{(2)}, x^{(3)}) \in \mathbb{Z}^3 \colon -L+1\le x^{(i)}\le  L,\ i=1,2,3 \}
$$
with a positive integer $L$, and the periodic boundary condition.
The lattice $\Lambda$ can be considered as
the three-dimensional torus. We write $e_\mu$ for the unit vector whose $\mu$-th component is $1$.


We write $\Psi_\sigma(x)$ for the fermion operator with the spin $\sigma=\uparrow,\downarrow$ at the site $x\in \Lambda$.
Each $\Psi_\sigma(x)$ has two components as follows:
\begin{equation*}
	\Psi_\sigma(x)=\begin{pmatrix}
		\psi_\sigma^{(1)}(x) \\ \psi_\sigma^{(2)}(x)
	\end{pmatrix}.
\end{equation*}
The components of the operators obey the anti-commutation relations,
\begin{equation*}
	\Bigl\{\psi_\sigma^{(i)}(x),\bigl[\psi_{\sigma'}^{(j)}(y)\bigr]^\dagger\Bigr\}=\delta_{\sigma,\sigma'}\delta_{i,j}\delta_{x,y}
\end{equation*}
and
\begin{equation*}
	\Bigl\{\psi_\sigma^{(i)}(x),\psi_{\sigma'}^{(j)}(y)\Bigr\}=0
\end{equation*}
for $x,y\in\Lambda$. Further, we introduce three $2\times 2$ matrices,
\begin{equation*}
	\alpha_1=\begin{pmatrix}
		0 & 1 \\ 1 & 0
	\end{pmatrix},
	\quad
	\alpha_2=\begin{pmatrix}
		0 & -i \\ i & 0
	\end{pmatrix}
	\quad \mbox{and} \quad
	\alpha_3=\begin{pmatrix}
		1 & 0 \\ 0 & -1
	\end{pmatrix}.
\end{equation*}
These matrices $\alpha_i$ act on the two component vectors $\Psi_\sigma(x)$ for $\sigma=\uparrow,\downarrow$.
We also write
\begin{equation*}
	\Psi(x)=\begin{pmatrix}
		\Psi_\uparrow(x) \\ \Psi_\downarrow(x)
	\end{pmatrix}
	\quad \mbox{and} \quad
	\alpha_i\Psi(x)=\begin{pmatrix}
		\alpha_i\Psi_\uparrow(x) \\ \alpha_i\Psi_\downarrow(x)
	\end{pmatrix},
	\quad i=1,2,3,
\end{equation*}
for short.
Then the one-particle Hilbert space is $\ell^2(\Lambda)\otimes \mathbb{C}^2_{\text{spin}} \otimes \mathbb{C}^2$, and many-particle Hamiltonians act on the corresponding fermion Fock space.
	Here we identify $\alpha_j$ with $1_{\text{spin}}\otimes \alpha_j$ on $\mathbb{C}^2_{\text{spin}} \otimes \mathbb{C}^2$.

Our Hamiltonian consists of three terms as follows:
\begin{equation}
	\label{Ham}
	H^{(\Lambda)}(B)=H_{\rm K}^{(\Lambda)}+H_{\rm int}^{(\Lambda)}+H_{\rm SBF}^{(\Lambda)}(B).
\end{equation}
The first term in the right-hand side is the kinetic Hamiltonian of Weyl type given by
	\begin{equation}
		\begin{split}
			\label{HK}
			H_{\rm K}^{(\Lambda)}&:=it \sum_{x\in\Lambda\subset\mathbb{Z}^3}
			\Bigl\{[\Psi^\dagger(x)\alpha_1\Psi(x+e_1)-\Psi^\dagger(x+e_1)\alpha_1\Psi(x)]\\
			&\qquad \qquad +[\Psi^\dagger(x)\alpha_2\Psi(x+e_2)-\Psi^\dagger(x+e_2)\alpha_2\Psi(x)]\\
			&\qquad \qquad +[\Psi^\dagger(x)\alpha_3\Psi(x+e_3)-\Psi^\dagger(x+e_3)\alpha_3\Psi(x)]\Bigr\}\\
			&=it\sum_{x\in \Lambda}\sum_{\mu=1}^3\sum_{\sigma=\uparrow, \downarrow}\left[\Psi_\sigma^\dagger(x)\alpha_\mu\Psi_\sigma(x+e_\mu)-\Psi_\sigma^\dagger(x+e_\mu)\alpha_\mu\Psi_\sigma(x)\right]
		\end{split}
\end{equation}
with the hopping parameter $t\in\mathbb{R}$.
Here we use the $\alpha_j$ in each direction $e_j$. Otherwise our construction establishing reflection positivity does not work, because we rely on the relations $\alpha_1 \alpha_j\alpha_1 =-\alpha_j$ for $j=2,3$ in Sec.~\ref{sec:x1=1/2}.

The second term $H_{\rm int}^{(\Lambda)}$ is the Hamiltonian of
the exchange interaction given by
\begin{equation}
	\label{Hint}
	H_{\rm int}^{(\Lambda)}:=J\sum_{x\in\Lambda}\sum_{\mu=1}^3\bigl[S^{(1)}(x)S^{(1)}(x+e_\mu)+S^{(2)}(x)S^{(2)}(x+e_\mu)
	+S^{(3)}(x)S^{(3)}(x+e_\mu)\bigr]
\end{equation}
with the coupling constant $J>0$, where the spin operators are given by
	\begin{equation*}
		S^{(j)}(x):=\Psi^\dagger(x)\tau_j\Psi(x)
		=
		\sum_{a=1,2}\sum_{\sigma, \sigma'=\uparrow,\downarrow} \psi^{(a)}_\sigma(x)^\dagger[\tau_j]_{\sigma \sigma'}  \psi^{(a)}_{\sigma'}(x)
	\end{equation*}
	for $j=1,2,3$.
Here the Pauli matrices $\tau_j$ act on the spin degrees of freedom: we identify $\tau_j$ with $\tau_j\otimes 1$ on $\mathbb{C}^2_{\text{spin}}\otimes \mathbb{C}^2$. The explicit forms are given by
\begin{equation*}
	\tau_1=\begin{pmatrix}
		0 & 1 \\ 1 & 0
	\end{pmatrix},
	\quad
	\tau_2=\begin{pmatrix}
		0 & -i \\ i & 0
	\end{pmatrix}
	\quad\mbox{and}\quad
	\tau_3=\begin{pmatrix}
		1 & 0 \\ 0 & -1
	\end{pmatrix}.
\end{equation*}
The third term of the Hamiltonian $H^{(\Lambda)}(B)$ is the symmetry-breaking source given by
\begin{equation}
	\label{HamSBF}
	H_{\rm SBF}^{(\Lambda)}(B):=-BO_\Lambda^{(1)}
\end{equation}
with the parameter $B\in\mathbb{R}$. Here $O_\Lambda^{(1)}$ denotes the order parameter
\begin{equation*}
	O_\Lambda^{(1)}:=\sum_{x\in\Lambda}(-1)^{x^{(1)}+x^{(2)}+x^{(3)}}S^{(1)}(x).
\end{equation*}
Our Hamiltonian conserves the total particle number and hence has a $\mathrm{U}(1)$ symmetry.
	At $B=0$, it also has a spin $\mathrm{SU}(2)$ symmetry, since the exchange interaction is isotropic in the spin space.

We stress that we have to impose the anti-periodic boundary condition~\cite{GK}
for the kinetic Hamiltonian (\ref{HK}), in order to realize the reflection positivity~\cite{JP}.

We write
\begin{equation}
	\label{Texpvalue}
	\langle \cdots \rangle_{\beta,B}^{(\Lambda)}:=\frac{1}{Z_{\beta,B}^{(\Lambda)}}{\rm Tr}\left[(\cdots)e^{-\beta H^{(\Lambda)}(B)}\right]
\end{equation}
for the thermal expectation value, where $Z_{\beta,B}^{(\Lambda)}={\rm Tr}\; e^{-\beta H^{(\Lambda)}(B)}$ is the partition function
with the inverse temperature $\beta$. We also write
\begin{equation*}
	\omega_B^{(\Lambda)}(\cdots):=\lim_{\beta\nearrow\infty}\langle \cdots \rangle_{\beta,B}^{(\Lambda)}
\end{equation*}
for the ground state expectation value.

In this paper, we prove
\begin{theorem}[Existence of LRO]
	\label{thm.main}
	Assume $B=0$.
	Then there exist positive constants $\alpha_0$, $\beta_0$, $c_0$ and an integer $L_0$ such that, whenever $L\ge L_0$, $|t|/J\le\alpha_0$ and $\beta\ge\beta_0$, it holds that
		\begin{equation}
			\label{theq.LRO}
			\frac{1}{|\Lambda|}\sqrt{\left\langle \left[O_\Lambda^{(1)}\right]^2\right\rangle_{\beta, 0}^{(\Lambda)}}\ge c_0>0,
		\end{equation}
		which shows the existence of long-range order.
		Moreover, the corresponding ground state bound follows by taking $\beta\nearrow\infty$.
\end{theorem}

\begin{remark}
	Applying the Koma--Tasaki theorem~\cite{KT} to this situation yields the existence of the corresponding spontaneous magnetization
	\[
	\lim_{B\searrow 0}\lim_{\Lambda\nearrow \mathbb{Z}^3}\frac{1}{|\Lambda|}\left\langle O_\Lambda^{(1)}\right\rangle_{\beta, B}^{(\Lambda)}>0.
	\]
	In other words, the equilibrium state exhibits the spontaneous symmetry breaking.
	The same holds true in the ground state, namely,
	\begin{equation*}
		\lim_{B\searrow 0}\lim_{\Lambda\nearrow \mathbb{Z}^3}\frac{1}{|\Lambda|}\omega_B^{(\Lambda)}(O_\Lambda^{(1)})>0.
	\end{equation*}
\end{remark}

	\begin{remark}
		At sufficiently high temperatures, we expect exponential decay of spin correlations and thus the absence of antiferromagnetic long-range order.
		The possible existence of intermediate phases remains to be determined.
\end{remark}

\section{Reflection positivity: Preliminary}
\label{sec:Reflec.Preli}

To begin with, we prepare some notations about reflection positivity in this section.
In particular, we will introduce Majorana fermions \cite{JP}, and real-valued functions
for the interaction Hamiltonian \cite{FSS,DLS}.

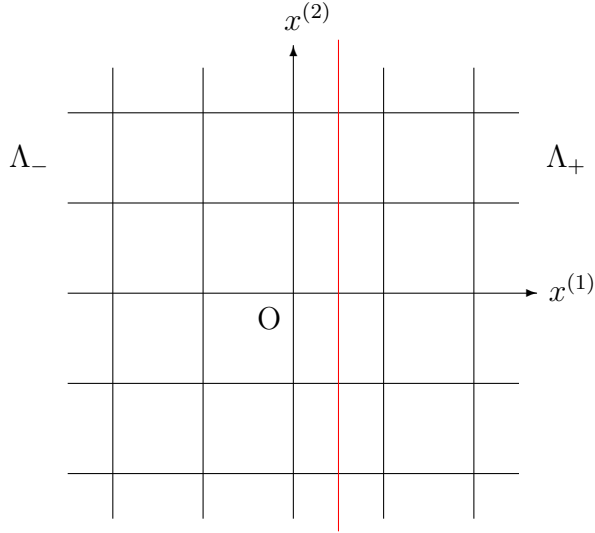
\begin{figure}[ht]
	\setlength{\unitlength}{1.2cm}
	\begin{center}
		\begin{picture}(5,5.6)(0,0)
			\put(0,0.5){\line(1,0){5}}
			\put(0,1.5){\line(1,0){5}}
			\put(0,2.5){\vector(1,0){5.2}}
			\put(0,3.5){\line(1,0){5}}
			\put(0,4.5){\line(1,0){5}}
			\put(0.5,0){\line(0,1){5}}
			\put(1.5,0){\line(0,1){5}}
			\put(2.5,0){\vector(0,1){5.25}}
			\put(3.5,0){\line(0,1){5}}
			\put(4.5,0){\line(0,1){5}}
			\put(5.33,2.4145){$x^{(1)}$}
			\put(2.1,2.1){O}
			\put(2.4,5.42){$x^{(2)}$}
			\put(3,-0.14){\color{red}{\line(0,1){5.45}}}
			\put(5.3,3.9){$\Lambda_+$}
			\put(-0.65,3.9){$\Lambda_-$}
		\end{picture}
	\end{center}
	\caption{The $x^{(1)}$-$x^{(2)}$ plane $(x^{(3)}=0)$ in the cubic lattice $\Lambda$ with the periodic boundary condition.
		The reflection plane $x^{(1)}=1/2$ is depicted by red color. By the two reflection planes $x^{(1)}=1/2, L+1/2$,
		the lattice $\Lambda$ is decomposed into
		the two sublattices $\Lambda_+$ and $\Lambda_-$.}
	\label{Lambda}
\end{figure}
Let $\Lambda' \subset \Lambda$ be a subset and $\mathcal{A}(\Lambda')$ be the algebra generated by
$\psi_\sigma^{(i)}(x)$ and $[\psi_{\sigma'}^{(j)}(y)]^\dagger$ for  $x, y \in \Lambda'$, $\sigma,\sigma'\in\{\uparrow,\downarrow\}$
and $i,j\in\{1,2\}$.
Since our $\Lambda$ is symmetric with respect to a plane with the periodic boundary condition,
there are a natural decomposition $\Lambda = \Lambda_- \cup \Lambda_+$ with $\Lambda_- \cap \Lambda_+ = \emptyset$
and a reflection map $r: \Lambda_{\pm} \to \Lambda_{\mp}$ satisfying $r(\Lambda_\pm) = \Lambda_\mp$,
as shown in Figure~\ref{Lambda}.
We write $\mathcal{A} = \mathcal{A}(\Lambda)$ and $\mathcal{A}_\pm = \mathcal{A}(\Lambda_\pm)$.
The reflection has an anti-linear representation $\vartheta: \mathcal{A}_\pm \to \mathcal{A}_\mp$ requiring \cite{JP}
\begin{align*}
	&\vartheta(\psi_\sigma^{(i)}(x)) = \psi_\sigma^{(i)}(r(x)), \quad
	\vartheta([\psi_\sigma^{(i)}(x)]^\dagger)= [\psi_\sigma^{(i)}(r(x))]^\dagger,\\
	&\vartheta(AB) = \vartheta(A)\vartheta(B), \quad \vartheta(A)^\dagger = \vartheta(A^\dagger) \quad \text{for } A, B \in \mathcal{A}.
\end{align*}

For $x \in \Lambda $, we introduce Majorana fermion operators $\xi_\sigma^{(i)}(x), \eta_\sigma^{(i)}(x)$ by
\begin{equation}
	\label{Majoranapsi}
	\xi_\sigma^{(i)}(x) := [\psi_\sigma^{(i)}(x)]^\dagger + \psi_\sigma^{(i)}(x), \quad
	\eta_\sigma^{(i)}(x) := i\{[\psi_\sigma^{(i)}(x)]^\dagger - \psi_\sigma^{(i)}(x)\},
\end{equation}
or equivalently,
\begin{equation}
	\label{psiMajorana}
	\psi_\sigma^{(i)}(x) = \frac{1}{2}[\xi_\sigma^{(i)}(x) + i\eta_\sigma^{(i)}(x)], \quad
	[\psi_\sigma^{(i)}(x)]^\dagger = \frac{1}{2}[\xi_\sigma^{(i)}(x) - i\eta_\sigma^{(i)}(x)].
\end{equation}
These satisfy $[\xi_\sigma^{(i)}(x)]^\dagger =\xi_\sigma^{(i)}(x)$,
$[\eta_\sigma^{(i)}(x)]^\dagger = \eta_\sigma^{(i)}(x)$, and the anti-commutation relations
\begin{align*}
	\{\xi_\sigma^{(i)}(x), \xi_{\sigma'}^{(j)}(y)\} &= 2\delta_{x, y}\delta_{i,j}\delta_{\sigma,\sigma'},  \quad
	\{{\eta}_\sigma^{(i)}(x), {\eta}_{\sigma'}^{(j)}(y)\} =2\delta_{x, y}\delta_{i,j}\delta_{\sigma,\sigma'}, \\
	\{\xi_\sigma^{(i)}(x), \eta_{\sigma'}^{(j)}(y)\} &=0.
\end{align*}

Next, following the idea of \cite{FSS}, we will introduce certain functions, $h_\mu$, on the lattice $\Lambda$, and
rewrite the interaction Hamiltonian $H_{\rm int}^{(\Lambda)}$ of (\ref{Hint}).
For this purpose, we note that
\begin{equation*}
	\sum_{x \in \Lambda}\sum_{\mu=1}^3S^{(i)}(x)S^{(i)}(x+e_\mu)
	=\frac{1}{2}\sum_{x\in\Lambda}\sum_{\mu=1}^3 [S^{(i)}(x)+S^{(i)}(x+e_\mu)]^2
	-3\sum_{x\in\Lambda}[S^{(i)}(x)]^2
\end{equation*}
for $i=1,3$, and
\begin{equation*}
	\sum_{x \in \Lambda}\sum_{\mu=1}^3S^{(2)}(x)S^{(2)}(x+e_\mu)=
	-\frac{1}{2}\sum_{x\in\Lambda}\sum_{\mu=1}^3 [S^{(2)}(x)-S^{(2)}(x+e_\mu)]^2
	+3\sum_{x\in\Lambda}[S^{(2)}(x)]^2.
\end{equation*}
Therefore, the interaction Hamiltonian $H_{\rm int}^{(\Lambda)}$ of (\ref{Hint}) can be written
\begin{align*}
	H_{\rm int}^{(\Lambda)}&=\frac{J}{2}\sum_{x \in \Lambda}\sum_{\mu=1}^3
	\bigl[S^{(1)}(x)+S^{(1)}(x+e_\mu)\bigr]^2
	+\frac{J}{2}\sum_{x \in \Lambda}\sum_{\mu=1}^3	\bigl[S^{(3)}(x)+S^{(3)}(x+e_\mu)\bigr]^2
	\\
	&\quad- \frac{J}{2}\sum_{x\in\Lambda}\sum_{\mu=1}^3[S^{(2)}(x)-S^{(2)}(x+e_\mu)]^2
	\\
	&\quad- 3J\sum_{x\in\Lambda}\bigl\{[S^{(1)}(x)]^2+[S^{(3)}(x)]^2-[S^{(2)}(x)]^2\bigr\}.
\end{align*}

Let $h_\mu(x)$ be a real-valued function on the lattice $\Lambda$ for $\mu=1,2,3$, and we introduce \cite{DLS,FSS}
\begin{equation}
	\begin{split}
		\label{Hinth}
		H_{\rm int}^{(\Lambda)}(h)&:=\frac{J}{2}\sum_{x \in \Lambda}\sum_{\mu=1}^3
		\bigl[S^{(1)}(x)+S^{(1)}(x+e_\mu)+h_\mu(x)\bigr]^2\\
		&\quad+\frac{J}{2}\sum_{x \in \Lambda}\sum_{\mu=1}^3	\bigl[S^{(3)}(x)+S^{(3)}(x+e_\mu)\bigr]^2
		- \frac{J}{2}\sum_{x\in\Lambda}\sum_{\mu=1}^3[S^{(2)}(x)-S^{(2)}(x+e_\mu)]^2
		\\
		&\quad- 3J\sum_{x\in\Lambda}\bigl\{[S^{(1)}(x)]^2+[S^{(3)}(x)]^2-[S^{(2)}(x)]^2\bigr\}.
	\end{split}
\end{equation}
We also write
\begin{equation}
	\label{HamBh}
	H^{(\Lambda)}(B,h):=H_{\rm K}^{(\Lambda)}+H_{\rm int}^{(\Lambda)}(h)+H_{\rm SBF}^{(\Lambda)}(B)
\end{equation}
for the whole Hamiltonian with the function $h$.

\section{Reflection with respect to the plane $x^{(1)}=1/2$}
\label{sec:x1=1/2}

Let us consider first the reflection with respect to the $x^{(1)}=1/2$ plane.\footnote{More precisely, perhaps we should say
	the reflection with respect to the two planes, $x^{(1)}=1/2,L+1/2$, because of the periodic boundary condition.}
(See Figure~\ref{Lambda}.) By this plane, we divide our finite lattice $\Lambda$ into two parts,
\[
\Lambda_- :=\{x \in \Lambda \colon -L+1 \le x^{(1)} \le 0\}
\quad \mbox{and}\quad
\Lambda_+ := \{x  \in \Lambda \colon 1 \le x^{(1)} \le L\}.
\]
In order to show that the present Hamiltonian has a reflection positivity with respect to this plane, we need some preparations.

\subsection{Three unitary transformations}

In the following, we will introduce three unitary transformations.
By using these unitary transformations, we can transform the Hamiltonian $H^{(\Lambda)}(B,h)$ of (\ref{HamBh})
to the desired form $\hat{H}^{(\Lambda)}(B,h)$ of (\ref{hatHBh}) below which shows reflection positivity
with respect to the above $x^{(1)}=1/2$ plane.

The first transformation is given by
\begin{equation*}
	\Psi(x)\rightarrow e^{i\pi x^{(2)}/2}\Psi(x).
\end{equation*}

Here the phase factor across the boundary bond from $x^{(2)}=L$ to $x^{(2)}=-L+1$ is
	$e^{i\pi(-L+1-L)/2}=i(-1)^L$.
	In the transformed hopping formulae below, the boundary condition in the $e_2$ direction is therefore understood to be anti-periodic for even $L$ and periodic for odd $L$.
We write $U_2$ for the corresponding unitary transformation on the fermion Fock space.

Let us consider first the kinetic Hamiltonian $H_{\rm K}^{(\Lambda)}$ of (\ref{HK}) under the above transformation $U_2$.
We write $H_{{\rm K}, \sigma}^{(\Lambda)}$ for the spin $\sigma\in \{\uparrow,\downarrow\}$ part of
the kinetic Hamiltonian $H_{\rm K}^{(\Lambda)}$ of (\ref{HK}). Then, by the transformation $U_2$,
the kinetic Hamiltonians $H_{{\rm K},\sigma}^{(\Lambda)}$ are transformed into the following forms:
\begin{equation}
	\begin{split}
		\label{tildeHKs}
		\tilde{H}_{{\rm K}, \sigma}^{(\Lambda)}&:=[U_2]^\dagger H_{{\rm K}, \sigma}^{(\Lambda)}U_2\\
		&=i t \sum_{x\in\Lambda\subset\mathbb{Z}^3}
		\Bigl\{[\Psi_\sigma^\dagger(x)\alpha_1\Psi_\sigma(x+e_1)-\Psi_\sigma^\dagger(x+e_1)\alpha_1\Psi_\sigma(x)]\\
		& \qquad \qquad +i[\Psi_\sigma^\dagger(x)\alpha_2\Psi_\sigma(x+e_2)+\Psi_\sigma^\dagger(x+e_2)\alpha_2\Psi_\sigma(x)]\\
		& \qquad \qquad +[\Psi_\sigma^\dagger(x)\alpha_3\Psi_\sigma(x+e_3)-\Psi_\sigma^\dagger(x+e_3)\alpha_3\Psi_\sigma(x)]\Bigr\}
	\end{split}
\end{equation}
for $\sigma\in\{\uparrow,\downarrow\}$.
When we use the real representation for the fermion field $\Psi(x)$, these right-hand sides become
pure imaginary hermitian by the expressions of the matrices, $\alpha_1, \alpha_2, \alpha_3$. This property
is crucial for the reflection positivity \cite{JP}.

In order to deal with the interaction Hamiltonian, we note that
\begin{equation*}
	\tilde{S}^{(i)}(x):=[U_2]^\dagger S^{(i)}(x)U_2=S^{(i)}(x) \quad \mbox{for \ }i=1,2,3.
\end{equation*}
Therefore,
\begin{equation*}
	\tilde{H}_{\rm int}^{(\Lambda)}(h):=[U_2]^\dagger H_{\rm int}^{(\Lambda)}(h)U_2
	=\tilde{H}_{\rm int,R}^{(\Lambda)}(h)+\tilde{H}_{{\rm int},{\rm I}}^{(\Lambda)},
\end{equation*}
where we have written
\begin{equation}
	\begin{split}
		\label{tildeHintRh}
		\tilde{H}_{\rm int,R}^{(\Lambda)}(h)&:=\frac{J}{2}\sum_{x \in \Lambda}\sum_{\mu=1}^3
		\Bigl\{\bigl[\tilde{S}^{(1)}(x)+\tilde{S}^{(1)}(x+e_\mu)+h_\mu(x)\bigr]^2\\
		&\quad +\bigl[\tilde{S}^{(3)}(x)+\tilde{S}^{(3)}(x+e_\mu)\bigr]^2\Bigr\}
		-3J\sum_{x\in\Lambda}\sum_{i=1,3}[\tilde{S}^{(i)}(x)]^2
	\end{split}
\end{equation}
and
\begin{equation}
	\label{tildeHintI}
	\tilde{H}_{{\rm int},{\rm I}}^{(\Lambda)}:=	-\frac{J}{2}\sum_{x\in\Lambda}\sum_{\mu=1}^3
	[\tilde{S}^{(2)}(x)-\tilde{S}^{(2)}(x+e_\mu)]^2
	+3J\sum_{x\in\Lambda}[\tilde{S}^{(2)}(x)]^2.
\end{equation}
The whole Hamiltonian $H^{(\Lambda)}(B,h)$ of (\ref{HamBh}) is transformed into
\begin{eqnarray}
	\label{tildeHmhf}
	\tilde{H}^{(\Lambda)}(B,h)&:=&[U_2]^\dagger H^{(\Lambda)}(B,h)U_2\nonumber\\
	&=&\tilde{H}_{{\rm K},\uparrow}^{(\Lambda)}+\tilde{H}_{{\rm K},\downarrow}^{(\Lambda)}
	+\tilde{H}_{\rm int,R}^{(\Lambda)}(h)+\tilde{H}_{\rm int, I}^{(\Lambda)}+\tilde{H}_{\rm SBF}^{(\Lambda)}(B),
\end{eqnarray}
where
\[
\tilde{H}_{\rm SBF}^{(\Lambda)}(B)
:=
U_2^\dagger H_{\rm SBF}^{(\Lambda)}(B)U_2
=
H_{\rm SBF}^{(\Lambda)}(B).
\]
We define the second unitary transformation $U(\alpha_1)$ as follows:
\begin{equation}
	\label{Ualpha1}
	[U(\alpha_1)]^\dagger \Psi_\sigma(x)U(\alpha_1)=\begin{cases}
		\alpha_1\Psi_\sigma(x) & \mbox{for \ } x\in\Lambda_+;\\
		\Psi_\sigma(x) & \mbox{for \ } x\in\Lambda_-
	\end{cases}
\end{equation}
for $\sigma\in\{\uparrow,\downarrow\}$.
This transformation $U(\alpha_1)$ can eliminate the matrix $\alpha_1$
at the bonds of the hopping Hamiltonian between the two sublattices $\Lambda_+$ and $\Lambda_-$.

In order to define the third unitary transformation $U_{\rm odd}$ of (\ref{Uodd}) below, we introduce \cite{FILS,GK}
\begin{equation}
	\label{uxsigma}
	u_{{\rm PH},\sigma}^{(i)}(x):=\left[\prod_{\substack{y\in\Lambda,\; j\in\{1,2\},\; {\sigma'}\in\{\uparrow,\downarrow\}
			: \\ (y,j,{\sigma'})\ne (x,i,\sigma)\; }}
	(-1)^{n_{\sigma'}^{(j)}(y)}\right]\bigl\{[\psi_\sigma^{(i)}(x)]^\dagger+\psi_\sigma^{(i)}(x)\bigr\},
\end{equation}
where we have written
\begin{equation*}
	n_{\sigma'}^{(j)}(y):=[\psi_{\sigma'}^{(j)}(y)]^\dagger\psi_{\sigma'}^{(j)}(y)
\end{equation*}
for $y\in\Lambda$ and ${\sigma'}\in\{\uparrow,\downarrow\}$. Then, one has
\begin{equation*}
	[u_{{\rm PH},\sigma}^{(i)}(x)]^\dagger \psi_{\sigma'}^{(j)}(y)u_{{\rm PH},\sigma}^{(i)}(x)=
	\begin{cases}
		[\psi_\sigma^{(i)}(x)]^\dagger, & \mbox{for}\; (y,j,{\sigma'})=(x,i,{\sigma}) ;\\
		\psi_{\sigma'}^{(j)}(y), & \mbox{otherwise}.
	\end{cases}
\end{equation*}
By using these operators, we define a particle-hole transformation on a sublattice by \cite{FILS,GK}
\begin{equation*}
	U_{{\rm odd},\sigma}:=\prod_{x\in\Lambda_{\rm odd}}\prod_{j\in\{1,2\}}u_{{\rm PH},\sigma}^{(j)}(x),
\end{equation*}
where we have written
\begin{equation*}
	\Lambda_{\rm odd}:=\{x\in\Lambda\colon x^{(1)}+x^{(2)}+x^{(3)}={\rm odd}\}.
\end{equation*}
For $i\in\{1,2\}$, one has
\begin{equation*}
	(U_{{\rm odd},\sigma})^\dagger \psi_\sigma^{(i)}(x)U_{{\rm odd},\sigma}=
	\begin{cases}
		[\psi_\sigma^{(i)}(x)]^\dagger & \mbox{for \ }x\in\Lambda_{\rm odd},\\
		\psi_\sigma^{(i)}(x) & \mbox{for \ }x\in\Lambda\backslash\Lambda_{\rm odd}.
	\end{cases}
\end{equation*}
In the case of $\sigma'\ne \sigma$, we have
\begin{equation*}
	(U_{{\rm odd},\sigma})^\dagger \psi_{\sigma'}^{(i)}(x)U_{{\rm odd},\sigma}=\psi_{\sigma'}^{(i)}(x)
\end{equation*}
for any $x\in\Lambda$ and $i=1,2$. Then, the third unitary transformation $U_{\rm odd}$ is defined by
\begin{equation}
	\label{Uodd}
	U_{\rm odd}:=U_{{\rm odd},\uparrow}U_{{\rm odd},\downarrow}.
\end{equation}
Clearly, one has
\begin{equation}
	\label{UPH}
	(U_{\rm odd})^\dagger \psi_\sigma^{(i)}(x)U_{\rm odd}=
	\begin{cases}
		[\psi_\sigma^{(i)}(x)]^\dagger & \mbox{for \ }x\in\Lambda_{\rm odd},\\
		\psi_\sigma^{(i)}(x) & \mbox{for \ }x\in\Lambda\backslash\Lambda_{\rm odd}.
	\end{cases}
\end{equation}

We write
\begin{equation*}
	\tilde{U}_1:=U(\alpha_1)U_{\rm odd}
\end{equation*}
for short, and
\begin{equation*}
	\hat{H}^{(\Lambda)}(B,h):=(\tilde{U}_1)^\dagger\tilde{H}^{(\Lambda)}(B,h)\tilde{U}_1
\end{equation*}
for the transformed Hamiltonian from (\ref{tildeHmhf}).
We want to decompose this Hamiltonian into three  parts \cite{GK} as follows.

	\begin{proposition}[Reflection positive representation]
		\label{prop:RP-x1}
		For any real-valued $h$, the transformed Hamiltonian
		$\hat{H}^{(\Lambda)}(B,h)$ admits the decomposition
		\begin{equation}
			\label{hatHBh}
			\hat{H}^{(\Lambda)}(B,h)
			=
			\hat{H}^+(B,h^+)
			+
			\hat{H}^-(B,h^-)
			+
			\hat{H}^0(h),
		\end{equation}
		where $\hat{H}^{\pm}(B,h^\pm)$ belong to
		$\mathcal{A}_{\pm}$, and
		\begin{equation}
			\label{hatH0h}
			\hat{H}^0(h)
			=
			\hat{H}_{\rm K}^0
			+
			\hat{H}_{\rm int,R}^0(h)
			+
			\hat{H}_{\rm int,I}^0.
		\end{equation}
		The three boundary terms are given in Lemmas~\ref{lem:RP-kinetic}--\ref{lem:RPim} below. Moreover,
		$
		\vartheta\bigl(\hat{H}^-(B,0)\bigr)=\hat{H}^+(B,0).
		$
\end{proposition}
Here, the two functions $h^\pm$ denote the restrictions of $h$ to $\Lambda_\pm$, and the rest term $\hat{H}^0(h)$ consists of certain operators
whose supports lie near the reflection plane. The reflections of the first two Hamiltonians satisfy
\begin{equation*}
	\vartheta(\hat{H}^+(B,h^+))=\hat{H}^-(B,\vartheta(h^+)),\quad
	\vartheta(\hat{H}^-(B,h^-))=\hat{H}^+(B,\vartheta(h^-)),
\end{equation*}
where $\vartheta(h^\pm)$ are the reflections of $h^\pm$.
Explicitly,
	\[
	(\vartheta h)_1(x)=-h_1(rx-e_1),\qquad
	(\vartheta h)_\mu(x)=-h_\mu(rx)\quad(\mu=2,3).
	\]

Once the above relations about the Hamiltonian $\hat{H}^{(\Lambda)}(B,h)$ of (\ref{hatHBh}) hold for all the reflection planes,
one can obtain the Gaussian domination bound (\ref{Zbound}) below in Sec.~\ref{GdomiInfrab} in the same way as in \cite{DLS,GK}.

We prove Proposition~\ref{prop:RP-x1} by treating each part of the Hamiltonian separately.

\subsection{Kinetic Hamiltonian}

Let us consider the kinetic Hamiltonians  (\ref{tildeHKs}).
	\begin{lemma}[Decomposition of the kinetic Hamiltonian]
		\label{lem:RP-kinetic}
		The transformed kinetic Hamiltonian admits the decomposition
		\begin{equation}
			\label{eq:RP-kinetic-decomposition}
			\hat{H}_{\rm K}^{(\Lambda)}
			:=
			\tilde{U}_1^\dagger U_2^\dagger
			H_{\rm K}^{(\Lambda)}
			U_2\tilde{U}_1
			=
			\hat{H}_{\rm K}^{-}
			+
			\hat{H}_{\rm K}^{+}
			+
			\hat{H}_{\rm K}^{0},
		\end{equation}
		where $\hat{H}_{\rm K}^{\pm} \in \mathcal{A}_{\pm}$ and $
		\vartheta(\hat{H}_{\rm K}^{-})
		=
		\hat{H}_{\rm K}^{+}$.
		The boundary term is
		\[
		\hat{H}_{\rm K}^{0}
		=
		\sum_{\sigma\in\{\uparrow,\downarrow\}}
		\hat{H}_{{\rm K},\sigma,1}^{0},
		\]
		where
		\begin{equation}
			\label{lemeq.hatHK}
			\begin{aligned}
				\hat{H}_{{\rm K},\sigma,1}^{0}
				&:=
				\tilde{U}_1^\dagger
				\tilde{H}_{{\rm K},\sigma,1}^{0}
				\tilde{U}_1
				\\
				&=
				\frac{it}{2}
				\sum_{\substack{x\in\Lambda\\x^{(1)}=0,-L+1}}
				\sum_{i=1}^{2}
				\Bigl[
				\xi_\sigma^{(i)}(x)
				\vartheta\bigl(\xi_\sigma^{(i)}(x)\bigr)
				+
				\eta_\sigma^{(i)}(x)
				\vartheta\bigl(\eta_\sigma^{(i)}(x)\bigr)
				\Bigr].
			\end{aligned}
		\end{equation}
	\end{lemma}

\begin{proof}
	We first decompose the kinetic Hamiltonian according to the hopping directions.
	\begin{equation*}
		\tilde{H}_{\rm K,\sigma}^{(\Lambda)}=\sum_{\mu=1}^3 \tilde{H}_{{\rm K,\sigma},\mu}^{(\Lambda)}
	\end{equation*}
	with
	\begin{equation*}
		\tilde{H}_{{\rm K,\sigma},\mu}^{(\Lambda)}
		:=i t\sum_{x\in\Lambda}[\Psi_{\rm \sigma}^\dagger(x)\alpha_\mu\Psi_{\rm \sigma}(x+e_\mu)
		-\Psi_{\rm \sigma}^\dagger(x+e_\mu)\alpha_\mu\Psi_{\rm \sigma}(x)]\quad \mbox{for \ } \mu=1,3,
	\end{equation*}
	and
	\begin{equation*}
		\tilde{H}_{{\rm K,\sigma},2}^{(\Lambda)}:=- t\sum_{x\in\Lambda}[\Psi_{\rm \sigma}^\dagger(x)\alpha_2\Psi_{\rm \sigma}(x+e_2)
		+\Psi_{\rm \sigma}^\dagger(x+e_2)\alpha_2\Psi_{\rm \sigma}(x)].
	\end{equation*}
	Further, the hopping Hamiltonians in the second and third directions can be decomposed into two parts as follows:
	\begin{equation*}
		\tilde{H}_{{\rm K,\sigma},2}^{(\Lambda)}=\tilde{H}_{{\rm K,\sigma},2}^++\tilde{H}_{{\rm K,\sigma},2}^-
	\end{equation*}
	with
	\begin{equation}
		\label{tildeHK2pm}
		\tilde{H}_{{\rm K,\sigma},2}^\pm:=- t\sum_{x\in\Lambda_\pm}[\Psi_{\rm \sigma}^\dagger(x)\alpha_2\Psi_{\rm \sigma}(x+e_2)
		+\Psi_{\rm \sigma}^\dagger(x+e_2)\alpha_2\Psi_{\rm \sigma}(x)]
	\end{equation}
	and
	\begin{equation*}
		\tilde{H}_{{\rm K,\sigma},3}^{(\Lambda)}=\tilde{H}_{{\rm K,\sigma},3}^++\tilde{H}_{{\rm K,\sigma},3}^-
	\end{equation*}
	with
	\begin{equation}
		\label{tildeHK3pm}
		\tilde{H}_{{\rm K,\sigma},3}^\pm:=i t\sum_{x\in\Lambda_\pm}[\Psi_{\rm \sigma}^\dagger(x)\alpha_3\Psi_{\rm \sigma}(x+e_3)
		-\Psi_{\rm \sigma}^\dagger(x+e_3)\alpha_3\Psi_{\rm \sigma}(x)].
	\end{equation}
	The kinetic term in the first direction is decomposed into three parts as follows:
	\begin{equation*}
		\tilde{H}_{{\rm K,\sigma},1}=\tilde{H}_{{\rm K,\sigma},1}^++\tilde{H}_{{\rm K,\sigma},1}^-+\tilde{H}_{{\rm K,\sigma},1}^0,
	\end{equation*}
	where
	\begin{equation}
		\label{tildeHK1+}
		\tilde{H}_{{\rm K,\sigma},1}^+:=i t\sum_{x\in\Lambda_+\; :\; x^{(1)}\ne L}
		[\Psi_{\rm \sigma}^\dagger(x)\alpha_1\Psi_{\rm \sigma}(x+e_1)-\Psi^\dagger_{\rm \sigma}(x+e_1)\alpha_1\Psi_{\rm \sigma}(x)],
	\end{equation}
	\begin{equation}
		\label{tildeHK1-}
		\tilde{H}_{{\rm K,\sigma},1}^-:=i t\sum_{x\in\Lambda_-\; :\; x^{(1)}\ne 0}
		[\Psi_{\rm \sigma}^\dagger(x)\alpha_1\Psi_{\rm \sigma}(x+e_1)-\Psi_{\rm \sigma}^\dagger(x+e_1)\alpha_1\Psi_{\rm \sigma}(x)],
	\end{equation}
	and
	\begin{equation}
		\begin{split}
			\label{tildeHK10}
			\tilde{H}_{{\rm K,\sigma},1}^0&:=
			i t\sum_{x\in\Lambda\; :\; x^{(1)}=0}[\Psi_{\rm \sigma}^\dagger(x)\alpha_1\Psi_\sigma(x+e_1)
			-\Psi_{\sigma}^\dagger(x+e_1)\alpha_1\Psi_{\sigma}(x)]\\
			&\quad+i t\sum_{x\in\Lambda\; :\: x^{(1)}=L}[\Psi_\sigma^\dagger(x_L^-)\alpha_1\Psi_\sigma(x)
			-\Psi_\sigma^\dagger(x)\alpha_1\Psi_\sigma(x_L^-)],
		\end{split}
	\end{equation}
	where $x_L^-:=(-L+1,x^{(2)},x^{(3)})$, and we have used the anti-periodic boundary condition for the kinetic terms.


	By using the unitary transformation $U(\alpha_1)$ of (\ref{Ualpha1}), we have
	\begin{align*}
		[U(\alpha_1)]^\dagger\tilde{H}_{{{\rm K},\sigma},1}^0U(\alpha_1)&=
		i t\sum_{x\in\Lambda\; :\; x^{(1)}=0}[\Psi_\sigma^\dagger(x)\Psi_\sigma(x+e_1)
		-\Psi_\sigma^\dagger(x+e_1)\Psi_\sigma(x)]\\
		&\quad+i t\sum_{x\in\Lambda\; :\: x^{(1)}=L}[\Psi_\sigma^\dagger(x_L^-)\Psi_\sigma(x)-\Psi_\sigma^\dagger(x)\Psi_\sigma(x_L^-)],
	\end{align*}
	where we have used $\alpha_1$ is self-adjoint, and $(\alpha_1)^2=1$.

	By using the Majorana fermions of (\ref{Majoranapsi}), one has
	\begin{equation*}
		[\psi_\sigma^{(i)}(x)]^\dagger \psi_\sigma^{(i)}(y)-[\psi_\sigma^{(i)}(y)]^\dagger\psi_\sigma^{(i)}(x)
		=\frac{1}{2}[\xi_\sigma^{(i)}(x)\xi_\sigma^{(i)}(y)+\eta_\sigma^{(i)}(x)\eta_\sigma^{(i)}(y)]\quad \mbox{for \ } x\ne y.
	\end{equation*}
	Therefore, we have
	\begin{equation}
		\begin{split}
			\label{Ualpha1tildeHK1}
			[U(\alpha_1)]^\dagger\tilde{H}_{{\rm K},\sigma,1}^0U(\alpha_1)&=
			\frac{i t}{2}\sum_{x\in\Lambda\; :\; x^{(1)}=0}\sum_{i=1}^2[\xi_\sigma^{(i)}(x)\xi_\sigma^{(i)}(x+e_1)
			+\eta_\sigma^{(i)}(x)\eta_\sigma^{(i)}(x+e_1)]\\
			&\quad+\frac{i t}{2}\sum_{x\in\Lambda\; :\: x^{(1)}=L}\sum_{i=1}^2
			[\xi_\sigma^{(i)}(x_L^-)\xi_\sigma^{(i)}(x)+\eta_\sigma^{(i)}(x_L^-)\eta_\sigma^{(i)}(x)].
		\end{split}
	\end{equation}
	From the representations (\ref{Majoranapsi}) of the Majorana fermions and (\ref{UPH}), one has
	\begin{equation*}
		(U_{\rm odd})^\dagger \xi_\sigma^{(i)}(x) U_{\rm odd}
		=\xi_\sigma^{(i)}(x)\quad \mbox{for \ } x\in\Lambda,
	\end{equation*}
	and
	\begin{equation*}
		(U_{\rm odd})^\dagger \eta_\sigma^{(i)}(x)U_{\rm odd}=
		\begin{cases}
			-\eta_\sigma^{(i)}(x) & \mbox{for \ } x\in\Lambda_{\rm odd};\\
			\eta_\sigma^{(i)}(x) & \mbox{for \ } x\in\Lambda\backslash\Lambda_{\rm odd}.
		\end{cases}
	\end{equation*}
	These observations imply that the transformation $U_{\rm odd}$
	changes the sign of the hopping amplitudes for $\eta_\sigma^{(i)}$.
	From this fact, $\tilde{U}_1=U(\alpha_1)U_{\rm odd}$, and the definition of the reflection map
	$\vartheta$ and (\ref{Ualpha1tildeHK1}), we obtain the desired expression,
	\begin{equation}
		\begin{split}
			\label{hatHK10}
			\hat{H}_{{\rm K,\sigma},1}^0&=
			\frac{i t}{2}\sum_{x\in\Lambda\; :\; x^{(1)}=0}\sum_{i=1}^2[\xi_\sigma^{(i)}(x)\xi_\sigma^{(i)}(x+e_1)
			-\eta_\sigma^{(i)}(x)\eta_\sigma^{(i)}(x+e_1)]\\
			&\quad+\frac{i t}{2}\sum_{x\in\Lambda\; :\: x^{(1)}=L}\sum_{i=1}^2[\xi_\sigma^{(i)}(x_L^-)\xi_\sigma^{(i)}(x)
			-\eta_\sigma^{(i)}(x_L^-)\eta_\sigma^{(i)}(x)]\\
			&=\frac{i t}{2}\sum_{x\in\Lambda\; :\; x^{(1)}=0}\sum_{i=1}^2[\xi_\sigma^{(i)}(x)\vartheta(\xi_\sigma^{(i)}(x))
			+\eta_\sigma^{(i)}(x)\vartheta(\eta_\sigma^{(i)}(x))]\\
			&\quad+\frac{i t}{2}\sum_{x\in\Lambda\; :\: x^{(1)}=-L+1}\sum_{i=1}^2[\xi_\sigma^{(i)}(x)\vartheta(\xi_\sigma^{(i)}(x))
			+\eta_\sigma^{(i)}(x)\vartheta(\eta_\sigma^{(i)}(x))].
		\end{split}
	\end{equation}
	This proves (\ref{lemeq.hatHK}), which has the required form for reflection positivity \cite{JP,GK}.

	As to $\tilde{H}_{{\rm K,\sigma},1}^\pm$ of (\ref{tildeHK1+}) and (\ref{tildeHK1-}), one notices that
	both of the two Hamiltonians do not change under the $U(\alpha_1)$ transformation of (\ref{Ualpha1}).
	Further, since the matrix $\alpha_1$ is symmetric, i.e., its transpose equals itself, one has
	\begin{align*}
		& (U_{\rm odd})^\dagger[\Psi_\sigma^\dagger(x)\alpha_1\Psi_\sigma(x+e_1)
		-\Psi_\sigma^\dagger(x+e_1)\alpha_1\Psi_\sigma(x)]U_{\rm odd}\\
		&=\Psi_\sigma^\dagger(x)\alpha_1\T\Psi_\sigma^\dagger(x+e_1)-\T\Psi_\sigma(x+e_1)\alpha_1\Psi_\sigma(x)
	\end{align*}
	for any $x\in\Lambda$, where the superscript `$\mathrm{t}$' denotes the transpose, namely
	\[
	\T\Psi^\dagger_\sigma(x)=
	\begin{pmatrix}
		\psi_\sigma^{(1)}(x)^\dagger \\ \psi_\sigma^{(2)}(x)^\dagger
	\end{pmatrix},
	\quad
	\T\Psi_\sigma(x)=\left(\psi_\sigma^{(1)}(x), \psi_\sigma^{(2)}(x) \right).
	\]
	Therefore, we have
	\begin{equation*}
		\hat{H}_{{\rm K,\sigma},1}^+:=(\tilde{U}_1)^\dagger\tilde{H}_{{\rm K, \sigma},1}^+\tilde{U}_1
		=i t\sum_{x\in\Lambda_+\; : \; x^{(1)}\ne L}[\Psi_\sigma^\dagger(x)\alpha_1\T\Psi_\sigma^\dagger(x+e_1)
		-\T\Psi_\sigma(x+e_1)\alpha_1\Psi_\sigma(x)]
	\end{equation*}
	and
	\begin{equation*}
		\hat{H}_{{\rm K,\sigma},1}^-:=(\tilde{U}_1)^\dagger\tilde{H}_{{\rm K,\sigma},1}^-\tilde{U}_1
		=i t\sum_{x\in\Lambda_-\; : \; x^{(1)}\ne 0}[\Psi_\sigma^\dagger(x)\alpha_1\T\Psi_\sigma^\dagger(x+e_1)
		-\T\Psi_\sigma(x+e_1)\alpha_1\Psi_\sigma(x)].
	\end{equation*}
	In addition, since the reflection map $\vartheta$ changes the hopping direction, we obtain
	\begin{equation*}
		\hat{H}_{{\rm K,\sigma},1}^+=\vartheta(\hat{H}_{{\rm K,\sigma},1}^-),
	\end{equation*}
	where we have used that the matrix $\alpha_1$ is real hermitian.

	Next, consider the kinetic Hamiltonian $\tilde{H}_{{\rm K,\sigma},3}^\pm$ of (\ref{tildeHK3pm}) in the third direction.
	From the properties of the matrices $\alpha_i$, one notices that
	\begin{equation*}
		[U(\alpha_1)]^\dagger \tilde{H}_{{\rm K,\sigma},3}^+U(\alpha_1)=-\tilde{H}_{{\rm K,\sigma},3}^+
		\quad \mbox{and} \quad
		[U(\alpha_1)]^\dagger \tilde{H}_{{\rm K,\sigma},3}^-U(\alpha_1)=\tilde{H}_{{\rm K,\sigma},3}^-.
	\end{equation*}
	Further, one has
	\begin{eqnarray*}
		& &(U_{\rm odd})^\dagger[\Psi_\sigma^\dagger(x)\alpha_3\Psi_\sigma(x+e_3)
		-\Psi_\sigma^\dagger(x+e_3)\alpha_3\Psi_\sigma(x)]U_{\rm odd}\\
		&=&\Psi_\sigma^\dagger(x)\alpha_3\T\Psi_\sigma^\dagger(x+e_3)-\T\Psi_\sigma(x+e_3)\alpha_3\Psi_\sigma(x)
	\end{eqnarray*}
	because the matrix $\alpha_3$ is symmetric. From these observations, we have
	\begin{equation}
		\label{hatHK3pm}
		\hat{H}_{{\rm K,\sigma},3}^\pm :=(\tilde{U}_1)^\dagger\tilde{H}_{{\rm K,\sigma},3}^\pm \tilde{U}_1
		=\mp i t\sum_{x\in\Lambda_\pm}[\Psi_\sigma^\dagger(x)\alpha_3\T\Psi_\sigma^\dagger(x+e_3)
		-\T\Psi_\sigma(x+e_3)\alpha_3\Psi_\sigma(x)]
	\end{equation}
	for $\sigma=\uparrow,\downarrow$. This implies
	\begin{equation*}
		\hat{H}_{{\rm K,\sigma},3}^+=\vartheta(\hat{H}_{{\rm K,\sigma},3}^-)\quad \mbox{for \ } \sigma=\uparrow,\downarrow
	\end{equation*}
	because the matrix $\alpha_3$ is real hermitian.

	Finally, let us consider $\tilde{H}_{{\rm K,\sigma},2}^\pm$ of (\ref{tildeHK2pm}). In the same way, one has
	\begin{equation*}
		[U(\alpha_1)]^\dagger\tilde{H}_{{\rm K,\sigma},2}^\pm U(\alpha_1)
		=\pm  t\sum_{x\in\Lambda_\pm}[\Psi_\sigma^\dagger(x)\alpha_2\Psi_\sigma(x+e_2)
		+\Psi_\sigma^\dagger(x+e_2)\alpha_2\Psi_\sigma(x)].
	\end{equation*}
	Note that
	\begin{align*}
		& (U_{\rm odd})^\dagger[\Psi_\sigma^\dagger(x)\alpha_2\Psi_\sigma(x+e_2)
		+\Psi_\sigma^\dagger(x+e_2)\alpha_2\Psi_\sigma(x)]U_{\rm odd}\\
		&=\Psi_\sigma^\dagger(x)\alpha_2\T\Psi_\sigma^\dagger(x+e_2)+\T\Psi_\sigma(x+e_2)\alpha_2\Psi_\sigma(x)
	\end{align*}
	for any $x\in\Lambda$, where we have used that the matrix $\alpha_2$ is anti-symmetric.
	By combining these three equations, we obtain
	\begin{equation*}
		\hat{H}_{{\rm K,\sigma},2}^\pm:=(\tilde{U}_1)^\dagger\tilde{H}_{{\rm K,\sigma},2}^\pm \tilde{U}_1
		=\pm  t\sum_{x\in\Lambda_\pm}[\Psi_\sigma^\dagger(x)\alpha_2\T\Psi_\sigma^\dagger(x+e_2)
		+\T\Psi_\sigma(x+e_2)\alpha_2\Psi_\sigma(x)]
	\end{equation*}
	for $\sigma=\uparrow,\downarrow$. Since the matrix $\alpha_2$ is pure imaginary hermitian, we have
	\begin{equation*}
		\hat{H}_{{\rm K,\sigma},2}^+=\vartheta(\hat{H}_{{\rm K,\sigma},2}^-)\quad \mbox{for \ } \sigma=\uparrow,\downarrow.
	\end{equation*}
	Taking
		\[
		\hat{H}_{\rm K}^{\pm}
		:=
		\sum_{\sigma\in\{\uparrow,\downarrow\}}
		\sum_{\mu=1}^{3}
		\hat{H}_{{\rm K},\sigma,\mu}^{\pm},
		\]
		we have the desired results.
\end{proof}

\subsection{Interaction Hamiltonian $\tilde{H}_{\rm int,R}^{(\Lambda)}(h)$}

Next, we consider the interaction Hamiltonian $\tilde{H}_{\rm int,R}^{(\Lambda)}(h)$ of (\ref{tildeHintRh}).

	\begin{lemma}[Transformation of the real interaction terms]
		\label{lem:RP-real}
		For $j=1,3$ we have
		\begin{align}
			\label{lemeq.S}
			\tilde{U_1}^\dagger U_2^\dagger S^{(j)}(x) U_2 \tilde{U_1}
			&=(-1)^{x^{(1)}+x^{(2)}+x^{(3)}}S^{(j)}(x),\\
			\vartheta(S^{(j)}(x))&=S^{(j)}(rx).
		\end{align}
		Moreover, it holds that
		\begin{equation}
			\label{eq:RP-real-interaction}
			\begin{aligned}
				\hat{H}_{\rm int,R}^{(\Lambda)}(h)
				&:=
				\tilde U_1^\dagger\tilde{H}_{\rm int,R}^{(\Lambda)}(h)\tilde U_1\\
				&=
				\frac{J}{2}\sum_{x\in\Lambda}\sum_{\mu=1}^{3}
				\Bigl\{
				\bigl[
				S^{(1)}(x)-S^{(1)}(x+e_\mu)
				+(-1)^{x^{(1)}+x^{(2)}+x^{(3)}} h_\mu(x)
				\bigr]^2
				\\
				&\hspace{35mm}
				+
				\bigl[
				S^{(3)}(x)-S^{(3)}(x+e_\mu)
				\bigr]^2
				\Bigr\}
				\\
				&\quad-
				3J\sum_{x\in\Lambda}\sum_{j=1,3}
				\bigl[S^{(j)}(x)\bigr]^2.
			\end{aligned}
		\end{equation}
		It admits the decomposition
		\begin{equation}
			\label{eq:RP-real-decomposition}
			\hat{H}_{\rm int,R}^{(\Lambda)}(h)
			=
			\hat{H}_{\rm int,R}^{-}(h)
			+
			\hat{H}_{\rm int,R}^{+}(h)
			+
			\hat{H}_{\rm int,R}^{0}(h),
		\end{equation}
		where the boundary term is
		\begin{equation}
			\label{eq:RP-real-boundary}
			\begin{aligned}
				\hat{H}_{\rm int,R}^{0}(h)
				=
				\frac{J}{2}
				\sum_{\substack{x\in\Lambda\\x^{(1)}=0,L}}
				\Bigl\{
				&
				\bigl[
				S^{(1)}(x)-S^{(1)}(x+e_1)
				+(-1)^{x^{(1)}+x^{(2)}+x^{(3)}}h_1(x)
				\bigr]^2
				\\
				&+
				\bigl[
				S^{(3)}(x)-S^{(3)}(x+e_1)
				\bigr]^2
				\Bigr\}.
			\end{aligned}
		\end{equation}
		In particular,
		\[
		\vartheta\bigl(\hat{H}_{\rm int,R}^{-}(0)\bigr)
		=
		\hat{H}_{\rm int,R}^{+}(0).
		\]
\end{lemma}
\begin{proof}
	For the part about the operator $\tilde{S}^{(1)}(x)$, we write
	\begin{equation}
		\label{tildeHtildeGamma1}
		\tilde{H}_{\tilde{S}^{(1)}}^{(\Lambda)}(h)=\sum_{\mu=1}^3\tilde{H}_{\tilde{S}^{(1)},\mu}^{(\Lambda)}(h)
		-3J\sum_{x\in\Lambda}[\tilde{S}^{(1)}(x)]^2,
	\end{equation}
	where
	\begin{equation*}
		\tilde{H}_{\tilde{S}^{(1)},\mu}^{(\Lambda)}(h):=\frac{J}{2}\sum_{x\in\Lambda}
		[\tilde{S}^{(1)}(x)+\tilde{S}^{(1)}(x+e_\mu)+h_\mu(x)]^2.
	\end{equation*}
	The Hamiltonian $\tilde{H}_{\tilde{S}^{(1)},1}^{(\Lambda)}(h)$ can be decomposed into three parts as follows:
	\begin{equation*}
		\tilde{H}_{\tilde{S}^{(1)},1}^{(\Lambda)}(h)
		=\tilde{H}_{\tilde{S}^{(1)},1}^+(h)+\tilde{H}_{\tilde{S}^{(1)},1}^-(h)+\tilde{H}_{\tilde{S}^{(1)},1}^0(h),
	\end{equation*}
	where
	\begin{equation*}
		\tilde{H}_{\tilde{S}^{(1)},1}^+(h):=\frac{J}{2}\sum_{\substack{x\in\Lambda_+:\\ x^{(1)}\ne L}}
		[\tilde{S}^{(1)}(x)+\tilde{S}^{(1)}(x+e_1)+h_1(x)]^2,
	\end{equation*}
	\begin{equation*}
		\tilde{H}_{\tilde{S}^{(1)},1}^-(h):=\frac{J}{2}\sum_{\substack{x\in\Lambda_-:\\ x^{(1)}\ne 0}}
		[\tilde{S}^{(1)}(x)+\tilde{S}^{(1)}(x+e_1)+h_1(x)]^2,
	\end{equation*}
	and
	\begin{equation}
		\label{Hint10h}
		\tilde{H}_{\tilde{S}^{(1)},1}^0(h):=\frac{J}{2}\sum_{\substack{x\in\Lambda:\\ x^{(1)}=0,L}}
		[\tilde{S}^{(1)}(x)+\tilde{S}^{(1)}(x+e_1)+h_1(x)]^2.
	\end{equation}
	Similarly, we have
	\begin{equation*}
		\tilde{H}_{\tilde{S}^{(1)},\mu}^{(\Lambda)}(h)
		=\tilde{H}_{\tilde{S}^{(1)},\mu}^+(h)+\tilde{H}_{\tilde{S}^{(1)},\mu}^-(h) \ \ \mbox{for \ } \mu=2,3,
	\end{equation*}
	where
	\begin{equation*}
		\tilde{H}_{\tilde{S}^{(1)},\mu}^\pm(h)
		:=\frac{J}{2}\sum_{x\in\Lambda_\pm}[\tilde{S}^{(1)}(x)+\tilde{S}^{(1)}(x+e_\mu)+h_\mu(x)]^2\quad \mbox{for \ }\mu=2,3.
	\end{equation*}
	From the definitions of $\tilde{S}^{(1)}(x)$ and $U(\alpha_1)$, one has
	\begin{equation*}
		[U(\alpha_1)]^\dagger \tilde{S}^{(1)}(x)U(\alpha_1)=\Psi_\uparrow^\dagger(x)\Psi_\downarrow(x)
		+\Psi_\downarrow^\dagger(x)\Psi_\uparrow(x).
	\end{equation*}
	Further, by using $\tilde{U}_1:=U(\alpha_1)U_{\rm odd}$, we have
	\begin{equation}
		\label{tildeU1tildeS1}
		(\tilde{U}_1)^\dagger \tilde{S}^{(1)}(x)\tilde{U}_1=(-1)^{x^{(1)}+x^{(2)}+x^{(3)}}
		[\Psi_\uparrow^\dagger(x)\Psi_\downarrow(x)+\Psi_\downarrow^\dagger(x)\Psi_\uparrow(x)].
	\end{equation}
	Therefore, for the Hamiltonian $\tilde{H}_{\tilde{S}^{(1)},1}^0(h)$ of (\ref{Hint10h}), we have
	\begin{equation}
		\label{hatH0S1}
		\tilde{U}_1^\dagger \tilde{H}_{\tilde{S}^{(1)},1}^0(h)\tilde{U}_1
		=\frac{J}{2}\sum_{\substack{x\in\Lambda:\\ x^{(1)}=0,L}}
		[\hat{S}^{(1)}(x)-\hat{S}^{(1)}(x+e_1)+(-1)^{x^{(1)}+x^{(2)}+x^{(3)}}h_1(x)]^2,
	\end{equation}
	where we have written
	\begin{equation*}
		\hat{S}^{(1)}(x):=\Psi_\uparrow^\dagger(x)\Psi_\downarrow(x)+\Psi_\downarrow^\dagger(x)\Psi_\uparrow(x).
	\end{equation*}
	This is the desired form \cite{DLS,GK} for getting a Gaussian domination.
	Similarly, one has
	\begin{equation*}
		\tilde{U}_1^\dagger \tilde{H}_{\tilde{S}^{(1)},1}^+(h)\tilde{U}_1
		=\frac{J}{2}\sum_{\substack{x\in\Lambda_+:\\ x^{(1)}\ne L}} [\hat{S}^{(1)}(x)-\hat{S}^{(1)}(x+e_1)
		+(-1)^{x^{(1)}+x^{(2)}+x^{(3)}}h_1(x)]^2
	\end{equation*}
	and
	\begin{equation*}
		\tilde{U}_1^\dagger \tilde{H}_{\tilde{S}^{(1)},1}^-(h)\tilde{U}_1
		=\frac{J}{2}\sum_{\substack{x\in\Lambda_-:\\ x^{(1)}\ne 0}} [\hat{S}^{(1)}(x)-\hat{S}^{(1)}(x+e_1)
		+(-1)^{x^{(1)}+x^{(2)}+x^{(3)}}h_1(x)]^2.
	\end{equation*}
	Moreover for $\mu=2,3$, we have
	\begin{equation*}
		\tilde{U}_1^\dagger \tilde{H}_{\tilde{S}^{(1)},\mu}^+(h)\tilde{U}_1
		=\frac{J}{2}\sum_{x\in\Lambda_+}[\tilde{S}^{(1)}(x)-\tilde{S}^{(1)}(x+e_\mu)+\tilde{h}_\mu(x)]^2
	\end{equation*}
	and
	\begin{equation*}
		\tilde{U}_1^\dagger \tilde{H}_{\tilde{S}^{(1)},\mu}^-(h)\tilde{U}_1
		=\frac{J}{2}\sum_{x\in\Lambda_-}[\tilde{S}^{(1)}(x)-\tilde{S}^{(1)}(x+e_\mu)+\tilde{h}_\mu(x)]^2
	\end{equation*}
	with $\tilde{h}_\mu(x):=(-1)^{x^{(1)}+x^{(2)}+x^{(3)}}h_\mu(x)$. In particular, when $h=0$, these imply
	\begin{equation*}
		\vartheta(\hat{H}_{\tilde{S}^{(1)},\mu}^-(0))=
		\hat{H}_{\tilde{S}^{(1)},\mu}^+(0)\quad \mbox{for \ } \mu=1,2,3,
	\end{equation*}
	where we have written
	\begin{equation*}
		\hat{H}_{\tilde{S}^{(1)},\mu}^\pm(h):=\tilde{U}_1^\dagger \tilde{H}_{\tilde{S}^{(1)},\mu}^\pm(h)\tilde{U}_1\quad
		\mbox{for \ }\mu=1,2,3.
	\end{equation*}
	Clearly, the second sum in the right hand side of (\ref{tildeHtildeGamma1}) can be treated in the same way.

	In the above argument, the relation (\ref{tildeU1tildeS1}) is crucial for the reflection positivity.
	Therefore, as to the operator $\tilde{S}^{(3)}(x)$ in the Hamiltonian $\tilde{H}_{\rm int,R}^{(\Lambda)}(h)$,
	it is enough to check the corresponding relation. Actually, in the same way, we have
	\begin{equation*}
		\tilde{U}_1^\dagger \tilde{S}^{(3)}(x)\tilde{U}_1=(-1)^{x^{(1)}+x^{(2)}+x^{(3)}}[\Psi_\uparrow^\dagger(x)\Psi_\uparrow(x)
		-\Psi_\downarrow^\dagger(x)\Psi_\downarrow(x)].
	\end{equation*}
	Substituting these identities into (\ref{tildeHintRh}) yields
		(\ref{eq:RP-real-interaction}). Together with the decompositions and reflection relations above, this completes the proof.
\end{proof}
\subsection{Interaction Hamiltonian $\tilde{H}_{\rm int,I}^{(\Lambda)}$}

Let us consider the interaction Hamiltonian $\tilde{H}_{{\rm int},{\rm I}}^{(\Lambda)}$ of (\ref{tildeHintI}).

	\begin{lemma}[Transformation of the imaginary interaction term]
		\label{lem:RPim}
		One has
		\begin{equation}
			\label{eq:RPimspin}
			\tilde{U_1}^\dagger U_2^\dagger S^{(2)}(x) U_2 \tilde{U_1}
			=
			S^{(2)}(x),
			\qquad
			\vartheta\bigl(S^{(2)}(x)\bigr)
			=
			-S^{(2)}(r(x)).
		\end{equation}
		Consequently, the transformed interaction is given by
		\begin{equation}
			\label{eq:RPimint}
			\hat{H}_{\rm int,I}^{(\Lambda)}
			:=
			\tilde{U}_1^\dagger
			\tilde{H}_{\rm int,I}^{(\Lambda)}
			\tilde{U}_1
			=
			-\frac{J}{2}
			\sum_{x\in\Lambda}\sum_{\mu=1}^{3}
			\bigl[S^{(2)}(x)-S^{(2)}(x+e_\mu)\bigr]^2
			+
			3J\sum_{x\in\Lambda}\bigl[S^{(2)}(x)\bigr]^2.
		\end{equation}
		It admits the decomposition
		\begin{equation}
			\label{eq:RPimdeco}
			\hat{H}_{\rm int,I}^{(\Lambda)}
			=
			\hat{H}_{\rm int,I}^{-}
			+
			\hat{H}_{\rm int,I}^{+}
			+
			\hat{H}_{\rm int,I}^{0},
		\end{equation}
		where
		\begin{equation}
			\label{eq:RPimbd}
			\hat{H}_{\rm int,I}^{0}
			=
			-\frac{J}{2}
			\sum_{\substack{x\in\Lambda\\x^{(1)}=0,L}}
			\bigl[S^{(2)}(x)-S^{(2)}(x+e_1)\bigr]^2,
		\end{equation}
		and
		\begin{equation}
			\label{eq:RPimpm}
			\vartheta\bigl(\hat{H}_{\rm int,I}^{-}\bigr)
			=
			\hat{H}_{\rm int,I}^{+}.
		\end{equation}
\end{lemma}
\begin{proof}
	It can be written
	\begin{equation}
		\label{HintI2}
		\tilde{H}_{{\rm int},{\rm I}}^{(\Lambda)}
		=\sum_{\mu=1}^3\tilde{H}_{\tilde{S}^{(2)},\mu}^{(\Lambda)}+3J\sum_{x\in\Lambda}[\tilde{S}^{(2)}(x)]^2,
	\end{equation}
	where
	\begin{equation*}
		\tilde{H}_{\tilde{S}^{(2)},\mu}^{(\Lambda)}
		:=-\frac{J}{2}\sum_{x\in\Lambda}[\tilde{S}^{(2)}(x)-\tilde{S}^{(2)}(x+e_\mu)]^2.
	\end{equation*}
	The Hamiltonian $\tilde{H}_{\tilde{S}^{(2)},1}^{(\Lambda)}$ can be decomposed into three parts as follows:
	\begin{equation*}
		\tilde{H}_{\tilde{S}^{(2)},1}^{(\Lambda)}
		=\tilde{H}_{\tilde{S}^{(2)},1}^++\tilde{H}_{\tilde{S}^{(2)},1}^-
		+\tilde{H}_{\tilde{S}^{(2)},1}^0,
	\end{equation*}
	where
	\begin{equation*}
		\tilde{H}_{\tilde{S}^{(2)},1}^+:=-\frac{J}{2}\sum_{\substack{x\in\Lambda_+:\\ x^{(1)}\ne L}}
		[\tilde{S}^{(2)}(x)-\tilde{S}^{(2)}(x+e_1)]^2,
	\end{equation*}
	\begin{equation*}
		\tilde{H}_{\tilde{S}^{(2)},1}^-:=-\frac{J}{2}\sum_{\substack{x\in\Lambda_-:\\ x^{(1)}\ne 0}}
		[\tilde{S}^{(2)}(x)-\tilde{S}^{(2)}(x+e_1)]^2,
	\end{equation*}
	and
	\begin{equation}
		\label{tildeHGamma210}
		\tilde{H}_{\tilde{S}^{(2)},1}^0:=-\frac{J}{2}\sum_{\substack{x\in\Lambda:\\ x^{(1)}=0,L}}
		[\tilde{S}^{(2)}(x)-\tilde{S}^{(2)}(x+e_1)]^2.
	\end{equation}
	Similarly, we have
	\begin{equation*}
		\tilde{H}_{\tilde{S}^{(2)},\mu}^{(\Lambda)}
		=\tilde{H}_{\tilde{S}^{(2)},\mu}^++\tilde{H}_{\tilde{S}^{(2)},\mu}^- \ \ \mbox{for \ } \mu=2,3,
	\end{equation*}
	where
	\begin{equation*}
		\tilde{H}_{\tilde{S}^{(2)},\mu}^\pm
		:=-\frac{J}{2}\sum_{x\in\Lambda_\pm}[\tilde{S}^{(2)}(x)-\tilde{S}^{(2)}(x+e_\mu)]^2\ \ \mbox{for \ }\mu=2,3.
	\end{equation*}

	One has
	\begin{equation*}
		[U(\alpha_1)]^\dagger \tilde{S}^{(2)}(x)U(\alpha_1)=
		-i[\Psi_\uparrow^\dagger(x)\Psi_\downarrow(x)-\Psi_\downarrow^\dagger(x)\Psi_\uparrow(x)].
	\end{equation*}
	Therefore, we obtain
	\begin{equation}
		\label{hatS2}
		\hat{S}^{(2)}(x):=(\tilde{U}_1)^\dagger \tilde{S}^{(2)}(x)\tilde{U}_1
		=-i[\Psi_\uparrow^\dagger(x)\Psi_\downarrow(x)-\Psi_\downarrow^\dagger(x)\Psi_\uparrow(x)].
	\end{equation}
	For the Hamiltonian $\tilde{H}_{\tilde{S}^{(2)},1}^0$ of (\ref{tildeHGamma210}), we have
	\begin{equation}
		\label{hatH0S2}
		\tilde{U}_1^\dagger \tilde{H}_{\tilde{S}^{(2)},1}^0\tilde{U}_1
		=-\frac{J}{2}\sum_{\substack{x\in\Lambda: \\ x^{(1)}=0,L}}[\hat{S}^{(2)}(x)-\hat{S}^{(2)}(x+e_1)]^2.
	\end{equation}
	This is the desired form \cite{DLS,GK} for getting a Gaussian domination
	because the right-hand side of (\ref{hatS2}) is pure imaginary. Similarly, one has
	\begin{equation*}
		\tilde{U}_1^\dagger \tilde{H}_{\tilde{S}^{(2)},1}^+\tilde{U}_1
		=-\frac{J}{2}\sum_{\substack{x\in\Lambda_+: \\ x^{(1)}\ne L}}[\hat{S}^{(2)}(x)-\hat{S}^{(2)}(x+e_1)]^2
	\end{equation*}
	and
	\begin{equation*}
		\tilde{U}_1^\dagger \tilde{H}_{\tilde{S}^{(2)},1}^-\tilde{U}_1
		=-\frac{J}{2}\sum_{\substack{x\in\Lambda_-: \\ x^{(1)}\ne 0}}[\hat{S}^{(2)}(x)-\hat{S}^{(2)}(x+e_1)]^2.
	\end{equation*}
	Further, we have
	\begin{equation*}
		\tilde{U}_1^\dagger \tilde{H}_{\tilde{S}^{(2)},\mu}^+\tilde{U}_1
		=-\frac{J}{2}\sum_{x\in\Lambda_+}[\hat{S}^{(2)}(x)-\hat{S}^{(2)}(x+e_\mu)]^2
	\end{equation*}
	and
	\begin{equation*}
		\tilde{U}_1^\dagger \tilde{H}_{\tilde{S}^{(2)},\mu}^-\tilde{U}_1
		=-\frac{J}{2}\sum_{x\in\Lambda_-}[\hat{S}^{(2)}(x)-\hat{S}^{(2)}(x+e_\mu)]^2
	\end{equation*}
	for $\mu=2,3$. From these observations, we have
	\begin{equation*}
		\vartheta\left(\hat{H}_{\hat{S}^{(2)},\mu}^-\right)=\hat{H}_{\hat{S}^{(2)},\mu}^+\quad \mbox{for \ } \mu=1,2,3,
	\end{equation*}
	where we have written
	\begin{equation*}
		\hat{H}_{\hat{S}^{(2)},\mu}^\pm:=\tilde{U}_1^\dagger \tilde{H}_{\tilde{S}^{(2)},\mu}^\pm\tilde{U}_1
		\quad \mbox{for \ } \mu=1,2,3,
	\end{equation*}
	and we have used that the operator $\hat{S}^{(2)}(x)$ is pure imaginary.

	Clearly, the second sum in the right-hand side of (\ref{HintI2}) can be treated in the same way.
	This completes the proof.
\end{proof}
\subsection{Hamiltonian $H_{\rm SBF}^{(\Lambda)}(B)$ of the symmetry-breaking field}

Finally, let us consider the Hamiltonian of the symmetry-breaking source, which is given by
\begin{equation*}
	H_{\rm SBF}^{(\Lambda)}(B)=-B\sum_{x\in\Lambda}(-1)^{x^{(1)}+x^{(2)}+x^{(3)}}S^{(1)}(x).
\end{equation*}
Clearly, this can be decomposed into two parts as follows:
\begin{equation*}
	H_{\rm SBF}^{(\Lambda)}(B)=H_{\rm SBF}^+(B)+H_{\rm SBF}^-(B)
\end{equation*}
with
\begin{equation*}
	H_{\rm SBF}^\pm(B):=-B\sum_{x\in\Lambda_\pm}(-1)^{x^{(1)}+x^{(2)}+x^{(3)}}S^{(1)}(x).
\end{equation*}
We write
\begin{equation*}
	\hat{H}_{\rm SBF}^\pm(B):=[U_2\tilde{U}_1]^\dagger H_{\rm SBF}^\pm(B)U_2\tilde{U}_1.
\end{equation*}
Then, from (\ref{tildeU1tildeS1}) and $\tilde{S}^{(1)}(x)=U_2^\dagger S^{(1)}(x)U_2$, we have
\begin{equation*}
	\vartheta(\hat{H}_{\rm SBF}^-(B))=\hat{H}_{\rm SBF}^+(B).
\end{equation*}
Together with the decompositions established above, this proves
	Proposition~\ref{prop:RP-x1}.
\section{Reflection with respect to the  plane $x^{(2)}=1/2$}
\label{sec:x2plane}

As to the reflection with respect to the $x^{(3)}=1/2$ plane,
the argument is the same as in the above case of the $x^{(1)}=1/2$ plane.
Therefore, it is enough to deal with the case of the reflection with respect to the $x^{(2)}=1/2$ plane.

We write
\begin{equation*}
	\mathcal{U}_3(\theta):=\exp\left[i\frac{\theta}{2}\alpha_3\right]
\end{equation*}
for the rotation by the generator $\alpha_3$ about the internal degrees of freedom with the angle $\theta\in[0,2\pi)$.
We also write $U_3(\theta)$ for the corresponding unitary operator on the fermion
Fock space, i.e.,
\begin{equation*}
	[U_3(\theta)]^\dagger\Psi_\sigma(x)U_3(\theta)=\mathcal{U}_3(\theta)\Psi_\sigma(x)
	\quad \mbox{for \ } \sigma\in\{\uparrow,\downarrow\} \ \mbox{and} \ x\in\Lambda.
\end{equation*}
Note that
\begin{equation*}
	[\mathcal{U}_3(\theta)]^\dagger\alpha_1\mathcal{U}_3(\theta)=\alpha_1\cos \theta +\alpha_2\sin\theta ,
\end{equation*}
\begin{equation}
	[\mathcal{U}_3(\theta)]^\dagger\alpha_2\mathcal{U}_3(\theta)=\alpha_2\cos \theta -\alpha_1\sin\theta ,
\end{equation}
and
\begin{equation*}
	[\mathcal{U}_3(\theta)]^\dagger\alpha_3\mathcal{U}_3(\theta)=\alpha_3.
\end{equation*}
Therefore, one has
\begin{equation}
	\label{rotalpha1}
	[\mathcal{U}_3(-\pi/2)]^\dagger\alpha_1\mathcal{U}_3(-\pi/2)=-\alpha_2,
\end{equation}
\begin{equation}
	\label{rotalpha2}
	[\mathcal{U}_3(-\pi/2)]^\dagger\alpha_2\mathcal{U}_3(-\pi/2)=\alpha_1,
\end{equation}
and
\begin{equation}
	\label{rotalpha3}
	[\mathcal{U}_3(-\pi/2)]^\dagger\alpha_3\mathcal{U}_3(-\pi/2)=\alpha_3.
\end{equation}
Clearly, these change only the matrices, $\alpha_1$ and $\alpha_2$, of
the hopping terms in the $x^{(1)}$ and $x^{(2)}$ directions in the kinetic Hamiltonian.
In addition, although the coefficient $ t$ in front of the matrix $\alpha_2$ changes its sign,
it does not affect the above argument about the reflection positivity in the case of the $x^{(1)}=1/2$ plane.

\section{Gaussian domination and infrared bound}
\label{GdomiInfrab}

Since we have been able to show that the Hamiltonian $\hat{H}^{(\Lambda)}(B,h)$ has the desired form of (\ref{hatHBh})
with $\hat{H}^0(h)$ of (\ref{hatH0h}), we can obtain the Gaussian domination bound
\begin{equation}
	\label{Zbound}
	{\rm Tr}\exp[-\beta H^{(\Lambda)}(B,h)]\le {\rm Tr}\exp[-\beta H^{(\Lambda)}(B,0)]
\end{equation}
for the Hamiltonian $H^{(\Lambda)}(B,h)$ of (\ref{HamBh}) in the same way as in \cite{GK}.
In fact, the decomposition form of (\ref{hatHBh}) with $\hat{H}^0(h)$ of (\ref{hatH0h}) is the same as~\cite[Eq.~(3.15)]{GK}.
Therefore, the proof is also the same as that of~\cite[Proposition~3.1]{GK}, and we omit the proof of the above bound.

By using the unitary transformation $U_{\rm odd}$ of (\ref{Uodd}), we write
\begin{equation*}
	H_\mathrm{odd}^{(\Lambda)}(B, h) := U_\mathrm{odd}^\dagger H^\mathrm{(\Lambda)}(B, h) U_\mathrm{odd}.
\end{equation*}
From the above bound (\ref{Zbound}), we obtain
\begin{equation}
	\label{Eq.Gaussian1}
	Z_\mathrm{odd}^{(\Lambda)}(B,h):=\mathrm{Tr}\left\{\exp[-\beta H_\mathrm{odd}^{(\Lambda)}(B, h)]\right\}
	\le Z_\mathrm{odd}^{(\Lambda)}(B,0).
\end{equation}
For any pairs of operators $\mathcal{A}$ and $\mathcal{B}$, we define the Duhamel two-point function by
\begin{align*}
	& (\mathcal{A},\mathcal{B})_\mathrm{odd}\\
	&\quad := \frac{1}{Z_\mathrm{odd}^{(\Lambda)}(B,0)}\int_0^1 ds\;
	\mathrm{Tr}\left\{ \exp[{-s \beta H_\mathrm{odd}^{(\Lambda)}(B, 0)}]\mathcal{A}
	\exp[{-(1-s)\beta H_\mathrm{odd}^{(\Lambda)}(B, 0)}] \mathcal{B} \right\}.
\end{align*}
We write
\begin{equation*}
	H_{{\rm int},1,\mu}^{(\Lambda)}(h):=\frac{J}{2}\sum_{x \in \Lambda} [S^{(1)}(x)+S^{(1)}(x+e_\mu) +h^{(\mu)}(x)]^2.
\end{equation*}
Note that
\begin{equation*}
	[U_\mathrm{odd}]^\dagger S^{(1)}(x) U_\mathrm{odd}= (-1)^{x^{(1)}+x^{(2)}+x^{(3)}} S^{(1)}(x).
\end{equation*}
Therefore, one has
\begin{align*}
	U_\mathrm{odd}^\dagger H_{\mathrm{int},1,\mu}^{(\Lambda)}(h) U_\mathrm{odd}
	&=
	\frac{J}{2} \sum_{x \in \Lambda} [S^{(1)}(x) - S^{(1)}(x+e_\mu) +\tilde{h}^{(\mu)}(x)]^2\\
	&=\frac{J}{2} \sum_{x \in \Lambda} \left[(S^{(1)}(x) - S^{(1)}(x+e_\mu))^2 + \tilde{h}^{(\mu)}(x)^2\right. \\
	&\quad \quad \quad \quad \quad \left. +2S^{(1)}(x) (\tilde{h}^{(\mu)}(x) - \tilde{h}^{(\mu)}(x-e_\mu))\right],
\end{align*}
where $\tilde{h}_\mu:=(-1)^{x^{(1)}+x^{(2)}+x^{(3)}}h_\mu$, and we have used
$$
\sum_{x \in \Lambda} [S^{(1)}(x) - S^{(1)}(x+e_\mu)]\tilde{h}^{(\mu)}(x)
= \sum_{x \in \Lambda}S^{(1)}(x) [\tilde{h}^{(\mu)}(x) - \tilde{h}^{(\mu)}(x-e_\mu)].
$$

Now we write
\begin{equation*}
	\partial_j \tilde{h}^{(\mu)} (x) := \tilde{h}^{(\mu)}(x) -\tilde{h}^{(\mu)}(x-e_j)\quad
	\mbox{and} \quad S^{(1)}[f] := \sum_x S^{(1)}(x)f(x)
\end{equation*}
for a complex-valued function $f$ on the lattice $\Lambda$.
Then we show the following inequality: For any complex-valued functions $\tilde{h}^{(\mu)}$,
\begin{equation}
	\label{Eq.infrared}
	\left(S^{(1)}\left[\overline{\sum_\mu \partial_\mu \tilde{h}^{(\mu)}}\right],
	S^{(1)}\left[\sum_\mu \partial_\mu \tilde{h}^{(\mu)}\right] \right)_{\rm odd}
	\le
	\frac{1}{\beta J} \sum_{\mu=1}^3 \sum_{x \in \Lambda} \vert \tilde{h}^{(\mu)}(x)\vert^2,
\end{equation}
where $\overline{z}$ denotes the complex conjugate of $z\in\mathbb{C}$.
Using $d^2 Z_\mathrm{odd}^{(\Lambda)}(B,\varepsilon h)/d\varepsilon^2 \vert_{\varepsilon = 0} \le 0$ by (\ref{Eq.Gaussian1})
and the identity followed from Duhamel's formula
\[
\left.\frac{d^2}{d \varepsilon^2} \mathrm{Tr} \left[\exp(-\beta H_\mathrm{odd}^{(\Lambda)}(B, 0)+\varepsilon\mathcal{A}) \right]
\right\vert_{\varepsilon=0}=
(\mathcal{A}, \mathcal{A})_{\rm odd} Z_\mathrm{odd}^{(\Lambda)}(B,0),
\]
we obtain (\ref{Eq.infrared}) for real-valued functions $h^{(\mu)}$.
With the help of the identity,
\begin{equation*}
	(\mathcal{A}^\dagger,\mathcal{A})_{\rm odd}= (\mathcal{A}_1,\mathcal{A}_1)_{\rm odd}
	+(\mathcal{A}_2,\mathcal{A}_2)_{\rm odd},
\end{equation*}
for $\mathcal{A}=\mathcal{A}_1+i\mathcal{A}_2$ with $\mathcal{A}_i^\dagger =\mathcal{A}_i$, $i=1, 2$,
the inequality (\ref{Eq.infrared}) holds for any complex-valued functions $\tilde{h}^{(\mu)}$.

We write $\Lambda^\ast$ for the dual lattice of $\Lambda$. Choosing
$$
\tilde{h}^{(\mu)}(x) = \vert\Lambda\vert^{-1/2} \{\exp[ip \cdot(x+ e_\mu)]- \exp[ip\cdot x]\}
$$
with $p= (p^{(1)},p^{(2)},p^{(3)})\in\Lambda^\ast$, we have
\[
\partial_\mu \tilde{h}^{(\mu)}(x) = -2\vert\Lambda\vert^{-1/2}e^{ip\cdot x}(1- \cos p^{(\mu)})
\]
and
\begin{equation}
	\label{Ep}
	\frac{1}{2}\sum_{x \in \Lambda} \sum_{\mu=1}^3 \vert \tilde{h}_\mu(x)\vert^2
	=\sum_{\mu=1}^3(1- \cos p^{(\mu)}) =: E_p.
\end{equation}
For $p \in \Lambda^\ast$,
let $\tilde S^{(1)}_p := \vert\Lambda\vert^{-1/2} \sum_x S^{(1)}(x) \exp[ip \cdot x]$.
By substituting these into the inequality (\ref{Eq.infrared}), we obtain the desired infrared bound,
\begin{equation}
	\label{Eq.IB}
	(\tilde S^{(1)}_p, \tilde S^{(1)}_{-p})_{\beta, B} \le \frac{1}{2 \beta J E_{p+Q}},
\end{equation}
where we have used $U_\mathrm{odd}^\dagger S^{(1)}(x) U_\mathrm{odd} = (-1)^{x^{(1)}+x^{(2)}+x^{(3)}} S^{(1)}(x)$,
$Q=(\pi, \dots, \pi)$ and the Duhamel two-point function for the Hamiltonian $H^{(\Lambda)}(B)$ in the right-hand side is given by
\begin{equation}
	\label{eq.Duhameldef}
	(\mathcal{A},\mathcal{B})_{\beta,B} := \frac{1}{Z_{\beta, B}^{(\Lambda)}}\int_0^1 ds\;
	\mathrm{Tr}\left[ e^{-s \beta H^{(\Lambda)}(B)} \mathcal{A} e^{-(1-s)\beta H^{(\Lambda)}(B) } \mathcal{B} \right].
\end{equation}

Let $C_p := \langle [\tilde S^{(1)}_{p}, [H^{(\Lambda)}(0), \tilde S^{(1)}_{-p}]] \rangle_{\beta, 0}^{(\Lambda)} $
be the expectation value (\ref{Texpvalue}) of the double commutator with $B=0$.
Then $C_p \ge 0$ follows by an eigenfunction expansion (see the next line of \cite[Eq.~(28)]{DLS}).
Applying~\cite[Thm.~3.2 \& Cor~3.2]{DLS} and the infrared bound (\ref{Eq.IB}), we have
\begin{equation}
	\label{eq.preLRO}
	\begin{split}
		\left\langle \tilde S^{(1)}_{p} \tilde S^{(1)}_{-p}+\tilde S^{(1)}_{-p} \tilde S^{(1)}_{p} \right\rangle_{\beta, 0}^{(\Lambda)}
		&\le
		\sqrt{\frac{C_p}{2JE_{p+Q}}} \coth\left(\sqrt{\frac{C_p \beta^2 J E_{p+Q} }{2 } }\right) \\
		&\le
		\sqrt{ \frac{C_p}{2 J E_{p+Q}}} + \frac{1}{\beta J E_{p+Q}},
	\end{split}
\end{equation}
where we have used the inequality $\coth x \le 1+ 1/x$.

In order to obtain a lower bound for LRO,
we want to use an upper bound for the expectation value of the interaction Hamiltonian, following \cite{KLS1}.
Actually, for the left-hand side of (\ref{eq.preLRO}), one has
\begin{equation}
	\label{eq.LHSLRO}
	\begin{split}
		&\sum_{p \in \Lambda^\ast} \left\langle \tilde S^{(1)}_{p}\tilde S^{(1)}_{-p} \right\rangle_{\beta, 0}^{(\Lambda)}\cos p^{(\mu)}
		\\
		&=
		\sum_{p \in \Lambda^\ast} \sum_{x, y \in \Lambda}
		\frac{\left\langle S^{(1)}(x) S^{(1)}(y) \right\rangle_{\beta, 0}^{(\Lambda)}}{ 2\vert\Lambda\vert} \left(e^{ip\cdot(x-(y-e_\mu))}
		+e^{ip\cdot((x-e_\mu) - y)}\right)\\
		&=\frac{1}{2\vert\Lambda\vert}
		\sum_{p \in \Lambda^\ast} \sum_{x, y \in \Lambda}e^{ip\cdot(x-y)}{\left\langle S^{(1)}(x) S^{(1)}(y+e_\mu)
			+ S^{(1)}(x+e_\mu) S^{(1)}(y)  \right\rangle_{\beta, 0}^{(\Lambda)}}\\
		&=
		\sum_{x \in \Lambda}\left\langle S^{(1)}(x) S^{(1)}(x+e_\mu)  \right\rangle_{\beta, 0}^{(\Lambda)}.
	\end{split}
\end{equation}
The rotational symmetry of SU(2) and the spatial symmetry
imply that the right-hand side is equal to the expectation value of
the interaction Hamiltonian divided by $9=3\times 3$ except for the coupling constant $J$. From the bound (\ref{eq.preLRO}), one has
\begin{equation}
	\label{Eq.LRO}
	\begin{split}
		&\frac{1}{ \vert\Lambda\vert} \sum_{p \in \Lambda^\ast}    \left\langle \tilde S^{(1)}_{p} \tilde S^{(1)}_{-p}
		+ \tilde S^{(1)}_{-p} \tilde S^{(1)}_{p} \right\rangle_{\beta, 0}^{(\Lambda)}
		\times \frac{1}{3}\sum_{\mu=1}^3 \left(-\cos p^{(\mu)}\right)\\
		&\le \vert\Lambda\vert^{-1}\sum_{p \neq Q}\left(  \frac{1}{  \beta J E_{p+Q}}
		+ \sqrt{\frac{C_p}{2J E_{p+Q}}}\right)\frac{1}{3} \left(-\sum_{\mu=1}^3\cos p^{(\mu)}\right)_+\\
		&\quad	+\frac{2}{\vert\Lambda\vert}{\left\langle \tilde S^{(1)}_{Q} \tilde S^{(1)}_{Q}  \right\rangle_{\beta,  0}^{(\Lambda)}},
	\end{split}
\end{equation}
where $F_+ := \max\{0, F\}$. Further, from (\ref{eq.LHSLRO}) and the spatial symmetry, the left-hand side is written
\begin{equation}
	\label{Eq.LROequi}
	\begin{split}
		&\frac{1}{ \vert\Lambda\vert} \sum_{p \in \Lambda^\ast}    \left\langle \tilde S^{(1)}_{p} \tilde S^{(1)}_{-p}
		+ \tilde S^{(1)}_{-p} \tilde S^{(1)}_{p} \right\rangle_{\beta, 0}^{(\Lambda)}
		\times \frac{1}{3}\sum_{\mu=1}^3 \left(-\cos p^{(\mu)}\right)\\
		&=-\frac{2}{ \vert\Lambda\vert} \sum_{x \in \Lambda}\left\langle S^{(1)}(x) S^{(1)}(x+e_\nu) \right\rangle_{\beta, 0}^{(\Lambda)}
	\end{split}
\end{equation}
for any $\nu=1,2,3$.

We next consider the last term in the right-hand side of (\ref{Eq.LRO}).
It can be written in terms of the long-range order parameter which is defined by
\begin{equation*}
	m_\mathrm{LRO}^{(\Lambda)}:=\frac{1}{|\Lambda|}\sqrt{\left\langle \left[O_\Lambda^{(1)}\right]^2\right\rangle_{\beta, 0}^{(\Lambda)}}
\end{equation*}
with
\begin{equation*}
	O_\Lambda^{(1)}=\sum_{x\in\Lambda}(-1)^{x^{(1)}+x^{(2)}+x^{(3)}}S^{(1)}(x).
\end{equation*}
Actually, the long-range order parameter can be written in the form
\begin{equation}
	\label{MLROS3Q}
	\begin{split}
		(m_\mathrm{LRO}^{(\Lambda)})^2
		&=
		\vert\Lambda\vert^{-2} \sum_{x, y \in \Lambda} (-1)^{x^{(1)}+x^{(2)}+x^{(3)}}(-1)^{y^{(1)}+y^{(2)}+y^{(3)}}
		\left\langle S^{(1)}(x)S^{(1)}(y) \right\rangle_{\beta,  0}^{(\Lambda)}\\
		&=
		\vert\Lambda\vert^{-1} \left\langle  \tilde S^{(1)}_{Q} \tilde S^{(1)}_{Q} \right\rangle_{\beta,  0}^{(\Lambda)},
	\end{split}
\end{equation}
where we have used $\tilde S^{(1)}_p := \vert\Lambda\vert^{-1/2} \sum_x S^{(1)}(x) \exp[ip \cdot x]$.

In order to estimate the first sum in the right-hand side of (\ref{Eq.LRO}), we need to evaluate the double commutator in $C_p$.
We first consider the free part $H_K^{(\Lambda)}$ of the present Hamiltonian $H^{(\Lambda)}(B)$ with the mass parameter $B=0$.
Using the commutation relations, one has
\begin{equation}
	\label{eq.doubleCK}
	\left\| \left[\tilde S^{(1)}_{p}, \left[H_K^{(\Lambda)}, \tilde S^{(1)}_{-p}\right]\right]\right\|\le\mathcal{C}_1\vert t\vert
\end{equation}
with a positive constant $\mathcal{C}_1$.


Next, we give a bound for the interaction part $H_\mathrm{int}^{(\Lambda)} = H_\mathrm{int}^{(\Lambda)}(0)$.
By a direct calculation, we have
\begin{equation*}
	\label{eq.Dint}
	\sum_{i=1}^3\left[S^{(i)}(x)S^{(i)}(y), S^{(1)}(x) \right]=-2iS^{(3)}(x)S^{(2)}(y)+2iS^{(2)}(x)S^{(3)}(y)
\end{equation*}
for $x \neq y$. Therefore, we have
\begin{equation*}
	\label{eq.Dinte}
	\left[ S^{(1)}(x), \sum_{i=1}^3\left[S^{(i)}(x)S^{(i)}(y), S^{(1)}(x) \right]\right]=
	-4 \left[S^{(2)}(x)S^{(2)}(y) +S^{(3)}(x)S^{(3)}(y) \right],
\end{equation*}
and
\begin{equation*}
	\label{eq.Dinter}
	\left[ S^{(1)}(y), \sum_{i=1}^3\left[S^{(i)}(x)S^{(i)}(y), S^{(1)}(x) \right]\right]=
	4\left[S^{(2)}(x)S^{(2)}(y) +S^{(3)}(x)S^{(3)}(y)\right].
\end{equation*}
Since $[S^{(i)}(x), S^{(j)}(y)] =0$ for $x \neq y$, one has
\begin{eqnarray}
	\label{eq.Dintera}
	& &\left[\tilde S^{(1)}_{p}, \left[H_\mathrm{int}^{(\Lambda)}, \tilde S^{(1)}_{-p}\right]\right]\nonumber\\
	&=&
	\frac{J}{\vert\Lambda\vert}\sum_{\mu=1}^3\sum_{x \in \Lambda}
	\Big[S^{(1)}(p;x,x+e_\mu),
	\sum_{i=1}^3\left[S^{(i)}(x)S^{(i)}(x+e_\mu), S^{(1)}(-p;x,x+e_\mu) \right]\Bigr]\nonumber\\
	&=&
	-\frac{8J}{\vert\Lambda\vert}\sum_{\mu=1}^3\left(1-\cos p^{(\mu)}\right)\sum_{x \in \Lambda}
	[S^{(2)}(x)S^{(2)}(x+e_\mu) +S^{(3)}(x)S^{(3)}(x+e_\mu)],\nonumber\\
\end{eqnarray}
where we have written
$$
S^{(1)}(p;x,x+e_\mu):=e^{ip\cdot x} S^{(1)}(x) +e^{ip\cdot (x+e_\mu)} S^{(1)}(x+e_\mu).
$$
Therefore, the spatial and the SU(2) symmetries of the Hamiltonian $H^{(\Lambda)}(0)$ with $B=0$ imply
\begin{equation}
	\label{eq.Dinteract}
	\begin{split}
		&\left\langle\left[\tilde S^{(1)}_{p}, \left[H_\mathrm{int}^{(\Lambda)}, \tilde S^{(1)}_{-p}\right]\right]
		\right\rangle_{\beta, 0}^{(\Lambda)}\\
		&=
		-\frac{16J}{\vert\Lambda\vert}\sum_{x \in \Lambda}
		\left\langle S^{(1)}(x)S^{(1)}(x+e_1)  \right\rangle_{\beta, 0}^{(\Lambda)} \sum_{\mu=1}^3\left(1-\cos p^{(\mu)}\right)\\
		&=-\frac{16JE_p}{\vert\Lambda\vert}\sum_{x \in \Lambda}
		\left\langle S^{(1)}(x)S^{(1)}(x+e_1)  \right\rangle_{\beta, 0}^{(\Lambda)},
	\end{split}
\end{equation}
where we have used the expression (\ref{Ep}) of $E_p$.
Combining (\ref{eq.doubleCK}) and (\ref{eq.Dinteract}) yields
\begin{align*}
	C_p&=\langle [\tilde{S}_p^{(1)},[H^{(\Lambda)}(0),\tilde{S}_{-p}^{(1)}]]\rangle_{\beta,0}^{(\Lambda)}\\
	&=\langle [\tilde{S}_p^{(1)},[H_K^{(\Lambda)},\tilde{S}_{-p}^{(1)}]]\rangle_{\beta,0}^{(\Lambda)}
	+\langle [\tilde{S}_p^{(1)},[H_{\rm int}^{(\Lambda)},\tilde{S}_{-p}^{(1)}]]\rangle_{\beta,0}^{(\Lambda)}\\
	&\le \mathcal{C}_1\vert t\vert +16JE_p\mathcal{E}_0^{(\Lambda)},
\end{align*}
where we have written
\[
\mathcal{E}_0^{(\Lambda)}:= -\vert\Lambda\vert^{-1}\sum_{x \in \Lambda}
\left\langle S^{(1)}(x)S^{(1)}(x+e_1)  \right\rangle_{\beta, 0}^{(\Lambda)}.
\]
We will show $\mathcal{E}_0^{(\Lambda)} \ge 0$ later. From this bound, one has
\begin{equation}
	\label{eq.Cpbound}
	\begin{split}
		\sqrt{\frac{C_p}{2  J E_{p+Q}}}&\le \sqrt{\frac{\mathcal{C}_1\vert t\vert+16JE_p\mathcal{E}_0^{(\Lambda)}}{2JE_{p+Q}}}\\
		&\le \sqrt{\frac{\mathcal{C}_1\vert t\vert}{2J E_{p+Q}}}
		+ {2\sqrt{2\mathcal{E}_0^{(\Lambda)}}} \sqrt{\frac{E_{p}}{E_{p+Q}}}.
	\end{split}
\end{equation}
Substituting (\ref{eq.Cpbound}) into (\ref{Eq.LRO}), we obtain
	\begin{equation}
		\label{eq.LROfinitevolume}
		\begin{split}
			\mathcal{E}_0^{(\Lambda)}
			&\le
			\frac{\mathcal{I}_3^{(\Lambda)}}{2\beta J}
			+\sqrt{\frac{\mathcal{C}_1|t|}{2J}}\,
			\mathcal{J}_3^{(\Lambda)}
			+\sqrt{2\mathcal{E}_0^{(\Lambda)}}\,
			\mathcal{K}_3^{(\Lambda)}
			+(m_{\mathrm{LRO}}^{(\Lambda)})^2,
		\end{split}
	\end{equation}
	where
	\begin{align*}
		\mathcal{I}_3^{(\Lambda)}
		&:=\frac{1}{|\Lambda|}
		\sum_{\substack{p\in\Lambda^\ast\\p\ne Q}}
		\frac{1}{E_{p+Q}},
		\qquad
		\mathcal{J}_3^{(\Lambda)}
		:=
		\frac{1}{|\Lambda|}\sum_{p\neq Q}\sqrt{\frac{1}{E_{p+Q}}},\\
		\mathcal{K}_3^{(\Lambda)}
		&:=\frac{1}{3|\Lambda|}\sum_{p\neq Q}
		\sqrt{\frac{E_p}{E_{p+Q}}}
		\left(-\sum_{\mu=1}^3\cos p^{(\mu)}\right)_+.
	\end{align*}
	Taking the infinite-volume limit, we obtain
\begin{equation}
	\label{eq.LROfin}
	\begin{split}
		\mathcal{E}_0
		&:=
		\lim_{\Lambda \nearrow \mathbb{Z}^3}\mathcal{E}_0^{(\Lambda)}
		\le \frac{\mathcal{I}_3}{2\beta J} + \sqrt{\frac{\mathcal{C}_1\vert t\vert}{2J}}\mathcal{J}_3+ \sqrt{2\mathcal{E}_0}\mathcal{K}_3
		+(m_\mathrm{LRO})^2,
	\end{split}
\end{equation}
where we have written
\begin{equation*}
	(m_\mathrm{LRO})^2=\lim_{\Lambda \nearrow \mathbb{Z}^3}(m^{(\Lambda)}_\mathrm{LRO})^2,
\end{equation*}
and the three constants, $\mathcal{I}_3$, $\mathcal{J}_3$, and $\mathcal{K}_3$, are given by
\begin{align*}
	\mathcal{I}_3&:= \frac{1}{(2\pi)^3}\int_{[-\pi, \pi]^3} \frac{dp}{E_p}, \quad
	\mathcal{J}_3:= \frac{1}{(2\pi)^3}\int_{[-\pi, \pi]^3} \frac{dp}{\sqrt{E_p}},\\
	\mathcal{K}_3&:=  \frac{1}{(2\pi)^3}\int_{[-\pi, \pi]^3} \; dp\; \frac{1}{3} \sqrt{\frac{E_p}{E_{p+Q}}}
	\left(-\sum_{\mu=1}^3\cos p^{(\mu)}\right)_+.
\end{align*}
Here, we have used (\ref{Eq.LROequi}), (\ref{MLROS3Q}) and
$$
0\le \frac{1}{3}\left(-\sum_{\mu=1}^3\cos p^{(\mu)}\right)_+\le 1.
$$
Since the three integrals, $\mathcal{I}_3$, $\mathcal{J}_3$ and $\mathcal{K}_3$, are all finite,
we can prove the existence of LRO,
i.e., $m_\mathrm{LRO}>0$ in the infinite-volume limit if $\mathcal{E}_0$ satisfies
\begin{equation}
	\label{eq.gsbound}
	\sqrt{\mathcal{E}_0} \left(\sqrt{\mathcal{E}_0}- \sqrt{2}\mathcal{K}_3\right) >0
\end{equation}
for $\beta J$ and $J/\vert t\vert$ both of which are sufficiently large .

The lower bound for $\mathcal{E}_0$ can be obtained as~\cite{GK1, FILS}.
By using $\|H_K^{(\Lambda)}\| \le \mathcal{C}_2\vert t\vert\vert\Lambda\vert$ with a positive constant $\mathcal{C}_2$
and the SU(2) symmetries, we  have the upper bound
\begin{equation}
	\label{eq.PHup}
	\left\langle  -H^{(\Lambda)}(0)  \right\rangle_{\beta, 0}
	\le \mathcal{C}_2\vert t\vert\vert\Lambda\vert
	-9J\sum_{x \in \Lambda} \left\langle S^{(3)}(x) S^{(3)}(x+e_1) \right\rangle_{\beta, 0}.
\end{equation}
To obtain the lower bound for $\mathcal{E}_0$, we use the following N\'eel state as a trial state:
\begin{equation*}
	\Phi:=\Biggl[\;\prod_{x \in \Lambda_\mathrm{odd}}[\psi_\uparrow^{(1)}]^\dagger(x)[\psi_\uparrow^{(2)}]^\dagger(x)\Biggr]
	\Biggl[\;\prod_{y \in \Lambda \backslash \Lambda_\mathrm{odd}}
	[\psi_\downarrow^{(1)}]^\dagger(y)[\psi_\downarrow^{(2)}]^\dagger(y)\Biggr]\vert 0\rangle,
\end{equation*}
where $\vert 0\rangle$ is the vacuum for fermions, namely $\psi_\sigma^{(i)}(x) \vert 0\rangle =0 $
for all $\sigma=\uparrow,\downarrow$, $i=1,2$ and $x \in \Lambda$.
We note that for any $x \in \Lambda$ and $\mu=1,2,3$
\begin{equation}
	\label{eq.Sev}
	\begin{split}
		\left\langle \Phi, S^{(i)}(x)S^{(i)}(x+e_\mu) \Phi\right\rangle
		=
		\begin{cases} 0, & (i=1,2);\\
			-4, & (i=3).
		\end{cases}
	\end{split}
\end{equation}
Next, we use the following inequality which is followed from the convexity (see, e.g., \cite[Proposition~2.5.4]{Ruelle}):
\begin{lemma}[Peierls's inequality]
	Let $A$ be a hermitian matrix and $\{\phi_i\}_i$ an orthonormal family.
	Then it holds that
	\[
	\sum_i \exp\left[-\langle \phi_i, A \phi_i\rangle\right]\le
	\mathrm{Tr} \exp(-A)
	\]
\end{lemma}
Using this, (\ref{eq.Sev}) and $\langle \Phi,H_K^{(\Lambda)}\Phi\rangle=0$ for the free part $H_K^{(\Lambda)}$ of
the present Hamiltonian, we obtain
\begin{align*}
	\mathrm{Tr} \,\exp\left[-\beta H^{(\Lambda)}(0)\right]
	&\ge
	\exp\left(\left\langle \Phi, -\beta H^{(\Lambda)}(0)\Phi \right\rangle\right)\\
	&\ge \exp\left[12\beta J\vert\Lambda\vert \right].
\end{align*}
Hence we have
\begin{equation}
	\label{eq.PHlow}
	\ln \mathrm{Tr} \, \exp\left[-\beta H^{(\Lambda)}(0)\right] \ge 12\beta J\vert\Lambda\vert.
\end{equation}
By the principle of maximum entropy for the Gibbs states (see, e.g.,~\cite[p.~90]{BR}), the following formula holds:
\[
\ln \mathrm{Tr} \, \exp\left[-\beta H^{(\Lambda)}(0)\right]
= \langle -\beta H^{(\Lambda)}(0)\rangle_{\beta, 0}^{(\Lambda)} - \mathrm{Tr} \left[\rho \ln \rho\right],
\]
where $\rho := e^{-\beta H^{(\Lambda)}(0)}/Z_{\beta, 0}^{(\Lambda)}$.
The concavity for the function $S: t \mapsto -t \ln t$ implies that for any $\sum_{j=1}^n \lambda _j=1$
\begin{align*}
	-\frac{1}{n} \sum_{j=1}^n\lambda_j \ln \lambda_j
	=
	\frac{1}{n}\sum_{j=1}^nS(\lambda_j)
	&\le
	S\left(\sum_{j=1}^n\frac{\lambda_j}{n} \right)
	=
	-\left(\frac{1}{n} \right) \ln \left(\frac{1}{n} \right)=\frac{1}{n}\ln(n),
\end{align*}
which yields
\[
-  \mathrm{Tr} \left[\rho \ln \rho\right]
\le \ln \mathrm{Tr}(1) =\ln 2^{4\vert\Lambda\vert}.
\]
Together with (\ref{eq.PHup}) and (\ref{eq.PHlow}), we arrive at
\begin{equation}
	\label{eq.fin}
	-9J\sum_{x \in \Lambda}\left\langle S^{(3)}(x) S^{(3)}(x+e_1) \right\rangle_{\beta,0}^{(\Lambda)}\ge
	\left(12J- \mathcal{C}_2 \vert t\vert\right) \vert \Lambda \vert - \frac{4}{\beta} \vert \Lambda \vert\ln 2.
\end{equation}
Therefore, by combining this with the SU(2) symmetry of the Hamiltonian $H^{(\Lambda)}(0)$, we have
\begin{equation}
	\label{eq.S1expv}
	-\frac{1}{\vert\Lambda\vert}\sum_{x \in \Lambda}\left\langle S^{(1)}(x) S^{(1)}(x+e_1)\right\rangle_{\beta,0}^{(\Lambda)}
	\ge
	\frac{4}{3}  -\frac{\mathcal{C}_2 \vert t\vert}{9J} -\frac{4}{\beta J}\ln 2.
\end{equation}
This shows
$$
\mathcal{E}_0^{(\Lambda)}=-\frac{1}{\vert\Lambda\vert}\sum_x\langle S^{(1)}(x) S^{(1)}(x+e_1) \rangle_{\beta, 0}^{(\Lambda)}
\ge \frac{4}{3}-\varepsilon
$$
with a small positive $\varepsilon$ which depends on $\vert t\vert/J$ and $1/(\beta J)$ both of which are sufficiently small.
Combining this with  (\ref{eq.LROfin}), LRO $m_\mathrm{LRO} >0$ exists
for large $\beta J$ and small enough $\vert t\vert/J$. Actually, one has
\begin{equation*}
	\sqrt{\frac{4}{3}}-\sqrt{2}\mathcal{K}_3=\sqrt{2}\left(\frac{\sqrt{6}}{3}-\mathcal{K}_3\right)>0
\end{equation*}
from the numerical values $\sqrt{6}/3 = 0.8164...$ and $K_3= 0.3498...$.

To conclude the proof, we note that the momentum sums in~\eqref{eq.LROfinitevolume} converge to the respective integrals as $L\to\infty$, independently of $\beta$ and $t$.
	Combining this with (\ref{eq.S1expv}) and the strict inequality above, we can take constants $\alpha_0,\beta_0,c_0>0$ and $L_0$ so that $m_{\mathrm{LRO}}^{(\Lambda)}\ge c_0$ whenever $L\ge L_0$, $\beta\ge\beta_0$ and $|t|/J\le\alpha_0$.
	This proves (\ref{theq.LRO}).

Finally, we prove the existence of LRO in the ground states.
	Taking $\beta\nearrow\infty$ in \eqref{eq.LROfinitevolume} before the infinite-volume limit, we obtain
	\[
	\sqrt{\mathcal{E}_\infty^{(\Lambda)}} \left(\sqrt{\mathcal{E}_\infty^{(\Lambda)}} -\sqrt{2}\mathcal{K}_3^{(\Lambda)}\right)
	\le
	\sqrt{\frac{\mathcal{C}_1\vert t\vert}{2J}}\mathcal{J}_3^{(\Lambda)}
	+\frac{1}{|\Lambda|}\omega_0^{(\Lambda)}\left(\tilde S^{(1)}_Q \tilde S^{(1)}_Q\right),
	\]
	where we have written
	\[
	\mathcal{E}_\infty^{(\Lambda)}
	:=
	-\frac{1}{|\Lambda|}\sum_{x\in \Lambda} \omega_0^{(\Lambda)} \left(S^{(1)} (x) S^{(1)} (x+e_1)\right).
	\]
	Using the lower bound (\ref{eq.S1expv}), we obtain LRO in the ground states
	\[
	\frac{1}{|\Lambda|}\omega_0^{(\Lambda)}\left(\tilde S^{(1)}_Q \tilde S^{(1)}_Q\right)\ge c_0^2,
	\]
	for $L\ge L_0$ and $|t|/J\le \alpha_0$ as above.
	This completes the proof of Theorem~\ref{thm.main}.

\section*{Acknowledgement}
We would like to thank Hosho Katsura and Hironobu Yoshida for helpful discussions and comments.
Y.~G. thanks the Yukawa Institute for Theoretical Physics at Kyoto University.
	Through discussions at the YITP workshop YITP-W-26-06 on ``Frontiers of Lattice Fermions,'' Y.~G. learned that the fermions we call `Weyl-type fermions' in this paper are known as `naive fermions' in lattice QCD.
Y.~G. is partially supported by JSPS Kakenhi Grant Number 23K12989.



\end{document}